\documentclass[12pt]{article}
\usepackage{epsfig}
\usepackage{axodraw}
\usepackage{amsmath}
\newcommand{\as}{\alpha_{\mathrm{s}}}
\newcommand{\aem}{\alpha_{\mathrm{em}}}
\renewcommand{\d}{\mathrm{d}}
\renewcommand{\u}{\mathrm{u}}
\newcommand{\s}{\mathrm{s}}
\renewcommand{\c}{\mathrm{c}}
\renewcommand{\b}{\mathrm{b}}
\newcommand{\X}{\mathbf{X}}

\newcommand{\e}{\mathrm{e}}

\newcommand{\g}{\mathrm{g}}
\newcommand{\p}{\mathrm{p}}
\newcommand{\q}{\mathrm{q}}

\newcommand{\qbar}{\mathrm{\overline{q}}}
\newcommand{\sbar}{\mathrm{\overline{s}}}
\newcommand{\cbar}{\mathrm{\overline{c}}}

\newcommand{\kT}{k_{\perp}}
\newcommand{\pT}{p_{\perp}}
\newcommand{\pTmin}{p_{\perp\mathrm{min}}}

\newcommand{\gast}{\gamma^*}
\newcommand{\ga}{\gamma}

\newcommand{\Py}{{\sc Pythia}}

\newcommand{\K}{\mathrm{K}}
\newcommand{\Jpsi}{\mathrm{J}/\psi}
\newcommand{\kTone}{k_{\perp 1}}
\newcommand{\kTtwo}{k_{\perp 2}}
\newcommand{\kTi}{k_{\perp i}}
\newcommand{\mr}{\mathrm}

\newcommand{\ra}{\rightarrow}

\newcommand{\lessim}{\raisebox{-0.8mm}%
{\hspace{1mm}$\stackrel{<}{\sim}$\hspace{1mm}}}
\def\Journal#1#2#3#4{{#1}{\bf #2} (#3) #4}

\def\NPB{{\it Nucl. Phys.~}{\bf B}}
\def\PLB{{\it Phys. Lett.~}{\bf  B}}
\def\JournalPLB#1#2#3{{\it Phys. Lett.~}{\bf {#1}B} (#2) #3}
\def\PL{{\it Phys. Lett.~}}
\def\PRL{\it Phys. Rev. Lett.~}
\def\PRD{{\it Phys. Rev.~}{\bf D}}
\def\ZPC{{\it Z. Phys.~}{\bf C}}

\def\JHEP{\it J. High Energy Phys.~}
\def\JPG{{\it J. Phys.~}{\bf G}}
\def\CPC{\it Computer Phys. Commun.~}
\def\PRP{\it Phys. Rept.~}
\def\PRV{\it Phys. Rev.~}
\def\EPJC{{\it Eur. Phys. J.~}{\bf C}}
\def\RMP{\it Rev. Mod. Phys.~}
\def\SJNP{\it Sov. J. Nucl. Phys.~}
\def\SPJP{\it Sov. Phys. JETP~}
\def\YF{\it Yad. Fiz.~}
\def\ZETF{\it Zh. Eksp. Teor. Fiz.~}
\newenvironment{Itemize}{\begin{list}{$\bullet$}%
{\setlength{\topsep}{0.2mm}\setlength{\partopsep}{0.2mm}%
\setlength{\itemsep}{0.2mm}\setlength{\parsep}{0.2mm}}}%
{\end{list}}
\newcounter{enumct}
\newenvironment{Enumerate}{\begin{list}{\arabic{enumct}.}%
{\usecounter{enumct}\setlength{\topsep}{0.2mm}%
\setlength{\partopsep}{0.2mm}\setlength{\itemsep}{0.2mm}%
\setlength{\parsep}{0.2mm}}}{\end{list}}
\newlength{\abstwidth}
\newlength{\captivewidth}
\newcommand{\captive}[1]{\rule{5mm}{0mm}%
\begin{minipage}{\captivewidth}%
\caption[small]{#1}\end{minipage}}
\begin{document}
%
 
\sloppy
 
\pagestyle{empty}
 
\begin{flushright}
LU TP 00--29\\
hep-ph/0007314\\
July 2000
\end{flushright}
 
\vspace{\fill}
 
\begin{center}
%
{\LARGE\bf Total Cross Sections and Event Properties\\[2mm]
from Real to Virtual Photons}\\[10mm]
{\Large 
Christer Friberg$^\star$ 
and Torbj\"orn Sj\"ostrand$^\star$
}\\[2mm]
{\it Department of Theoretical Physics,}\\[1mm]
{\it Lund University, Lund, Sweden}
\end{center}
 
\vspace{\fill}
\begin{center}
{\bf Abstract}\\[2ex]
\begin{minipage}{\abstwidth}
A model for total cross sections with virtual photons is presented. 
In particular $\gast\p$ and $\gast\gast$ cross sections are considered. 
Our approach extends on a model for photoproduction, where the total 
cross section is subdivided into three distinct event classes: 
direct, VMD and anomalous. 
With increasing photon virtuality, the latter two decrease in importance.
Instead Deep Inelastic Scattering dominates, with the direct class being 
the $\mathcal{O}(\alpha_s)$ correction thereof. 
Hence, the model provides a smooth transition between the 
two regions. By the breakdown into different event classes, one may
aim for a complete picture of all event properties. 
\end{minipage}
\end{center}

\vspace{\fill}

\footnoterule
{\footnotesize $^\star$christer@thep.lu.se, torbjorn@thep.lu.se}

\clearpage
\pagestyle{plain}
\setcounter{page}{1}

\section{Introduction}
\label{intro}

Traditionally, different descriptions are used for virtual and real
photons. Virtual photons in the DIS (Deeply Inelastic Scattering) 
region are normally described as devoid of any structure, while for 
the real ones, the possibility of hadronic-like fluctuations play an 
important r\^ole. In the region of intermediate $Q^2$, it should be 
possible to find a description starting from either extreme. Then the 
language may not always be unique, i.e. a given Feynman diagram may be 
classified in different ways. 

Several paths are possible; one is to have an explicit description in 
terms of higher-order Feynman diagrams~\cite{HOME}, which then may contain
several hard scales. However, in this article a main point is to obtain
a smooth transition to soft physics, beyond the region of validity of
perturbation theory, and then it is convenient to be able to consider 
transitions in one scale at a time.
In the following, we will therefore develop one specific approach, where 
the main principle is to characterize events by a set of standard scales,
such as the photon virtuality (or virtualities for $\ga\ga$), the
photon resolution scale(s), and the hard scale of a partonic subprocess.
Depending on the relative ordering of these scales, events are classified
in different categories. Special emphasis is normally put on the hardest 
scale of the event. This may determine e.g. whether an event is classified 
as a DIS or a resolved photon one. Matrix
elements are defined to lowest required order only, and higher-order 
corrections are approximated by parton showers.
For instance, if the hard scale is larger than the resolution scale of the 
photon, a partonic evolution is allowed between the two scales. 
The DGLAP equations are then suitable for this 
evolution~\cite{DGLAP} and are to be combined with the appropriate 
leading-order matrix elements, responsible for the hard scattering. 
In a $\ga\p$ event, the photon resolution scale may well be smaller than
both the photon virtuality and the hard-scattering scale.
In this way, partly unordered evolutions in $\pT$ are accounted for.
This is still less complete than allowed by the CCFM 
equations~\cite{CCFM,LDC}, where the ordering may be broken in several steps, 
but could well provide the bulk of non-ordering effects at current energies.

The classification used here is not an economical route, however, since it 
leads to many event classes. For studies e.g. of the total cross section in 
the intermediate-$Q^2$ region, it is cumbersome and not necessarily
better than existing approaches \cite{existingsigmagap}. However, by 
the breakdown into distinct event classes, the road is open to provide 
a (more or less) complete picture of all event properties. It is this 
latter aspect that has then guided the model development. 

As an example, the $\gast\p$ cross section is divided into a VMD, an 
anomalous, a direct and a DIS component. In the limit $Q^2 \rightarrow 0$, 
the DIS process $\gast\q\ra\q$ becomes kinematically forbidden and only
the first three event classes remain, reproducing
the result from the photoproduction model (except for some improvements). 
On the contrary, when $Q^2$ increases from zero to high values, the resolved 
processes decrease in importance (as given by dipole dampening factors), 
the direct ones also drops (by $Q^2$ dependence and a shrinking phase space) 
and finally only the DIS process remain. At intermediate $Q^2$ values, the 
direct processes and the DIS (+parton showers) process overlap, since, in 
some regions of phase space, they are equally valid descriptions of the same 
physics. It thus becomes necessary to avoid double-counting, e.g. by 
introducing Sudakov style form factors for the DIS process, suppressing those 
parton configurations covered by the direct processes. 

Based on the study of jet production by virtual photons~\cite{lutp9911} 
we extend the modeling to low--$\pT$ events. Clearly, a smooth transition 
from perturbative to non-perturbative physics is desired, and
 is achieved partly by allowing for multiple parton--parton scatterings.
These are needed to explain the underlying event activity seen in the data in
hadron--hadron collisions~\cite{multint}, and the ideas will be taken over to
the hadronic resolved photon components. At the same time, the multiple 
parton--parton interactions unitarize the jet cross sections, a necessity for
a decent growth of the total cross section with increasing energy. 

The outline of the paper is as follows. First, the different event classes are 
presented, including an extension to virtual photons of a model for 
photoproduction. How the event classes are combined, in order to 
avoid double-counting but still to cover the whole phase space, is described 
in detail. Then, results of the model are 
presented, with some comparisons to available data. In particular, total 
cross sections with virtual photons are shown with emphasis on the importance 
of the different event classes at various photon virtualities. Some 
distributions of event properties are shown to vary significantly with the 
photon virtuality, others less so, illustrating a complex model working 
smoothly between the regions of real and highly virtual photons. Similarities 
in $\ga\p$ and $\ga\ga$ events will be pointed out where appropriate. 
Finally, a summary and an outlook is given.

\section{Event Classes}
\label{sec:event}
 
In the following, we begin by a reminder on the models for DIS and
photoproduction, before embarking on the generalization also to  
intermediate virtualities in $\ga^*\p$ processes. The $\ga\ga$, 
$\ga^*\ga$ and $\ga^*\ga^*$ processes thereafter follow by 
an application of the same rules.

\subsection{Deeply Inelastic Scattering}
\label{sec:DIS}

\begin{figure}[t]
\begin{center}       
\begin{picture}(180,140)(10,60)
\ArrowLine(10,150)(80,150)\Text(35,160)[2]{\normalsize $\e (k)$}
\ArrowLine(80,150)(160,190)\Text(115,180)[10]{\normalsize $\e (k')$}
\Photon(80,150)(130,110){5}{6}
\Text(80,120)[]
{\normalsize $\gast (q)$}
\ArrowLine(10,90)(130,90)\Text(70,75)[]{\normalsize $\p (P)$}
\GOval(135,100)(20,10)(0){0.5}
\LongArrow(143,110)(180,130)
\LongArrow(144,105)(180,115)
\LongArrow(145,100)(180,100)\Text(190,100)[]{\normalsize $\X$}
\LongArrow(144,95)(180,85)
\LongArrow(143,90)(180,70)
\end{picture}     
\end{center}
\captive{
Deeply inelastic charged lepton-hadron scattering. (The four-momenta are 
given in parentheses.)
\label{fig:DIS}}
\end{figure}

The Deeply Inelastic Scattering of a high-energy charged lepton off 
a proton target, Fig.~\ref{fig:DIS}, involves a single electroweak boson 
exchange between a beam lepton and a target quark. 
At not too large $Q^2$ only photon exchange need be considered.
Then the double-differential $\e\p$ cross-section for DIS can be 
expressed in terms of the total cross-section 
for virtual transverse (T) and longitudinal (L) photons \cite{sigmaTL}: 
\begin{equation}
\frac{\d^2\sigma(\e\p \rightarrow \e\X)}{\d y \, \d Q^2}
=  
f_{\ga/\e}^{\mr{T}}(y,Q^2) \sigma_\mr{T}(y,Q^2) + 
f_{\ga/\e}^{\mr{L}}(y,Q^2) \sigma_\mr{L}(y,Q^2) ~,
\end{equation}
with the fluxes 
\begin{eqnarray}
f_{\ga/\e}^{\mr{T}}(y,Q^2) & = & \frac{\aem}{2\pi} 
\left( \frac{1+(1-y)^2}{y} \frac{1}{Q^2}-\frac{2m_{\e}^2y}{Q^4} \right)\;,\\
f_{\ga/\e}^{\mr{L}}(y,Q^2) & = & \frac{\aem}{2\pi} 
\frac{2(1-y)}{y} \frac{1}{Q^2}\;.
\label{LLogflux}
\end{eqnarray}
The conventional DIS variables
\begin{equation}
y=\frac{qP}{kP}\;,\hspace{4mm}
x=\frac{Q^2}{2qP}=\frac{Q^2}{Q^2+W^2-m_\p^2}\;,  \hspace{4mm}
Q^2=-q^2\;, \hspace{4mm}
W^2=(q+P)^2\;,
\label{DISvar}
\end{equation}
are related e.g. by
\begin{equation}
Q^2 = x y \, 2kP ~, \hspace{4mm} W^2 = (1-x) y \, 2kP + m_\p^2 ~,  
\end{equation}
where $2kP \approx (k+P)^2 = s$. Thus there are only two kinematical
degrees of freedom.

The cross-sections can be related to the proton structure functions 
$F_2$ and $F_\mr{L}$ by \cite{F2FLdef,Bauer}
\begin{equation}
F_2(x,Q^2)=
\frac{Q^2}{4\pi^2\aem}
\frac{(1-x)Q^2}{(Q^2+4m_\p^2x^2)}
(\sigma_\mr{T}+\sigma_{\mr{L}})\;,\hspace{2mm}
F_\mr{L}(x,Q^2)=
\frac{Q^2}{4\pi^2\aem}\frac{(1-x)Q^2}{(Q^2+4m_\p^2x^2)}
\sigma_\mr{L}
\end{equation}
 and the total virtual photon-proton cross section by
\begin{equation}
\sigma_\mr{tot}^{\gast\p}
\equiv\sigma_\mr{T}+\sigma_\mr{L}=
\frac{4\pi^2\aem}{Q^2}\frac{(Q^2+4m_\p^2x^2)}{(1-x)Q^2} F_2(x,Q^2)
\simeq\frac{4\pi^2\aem}{Q^2(1-x)} F_2(x,Q^2)\;.
\label{eq:F2}
\end{equation} 

In the parton model,
\begin{equation}
F_2(x,Q^2) = \sum_{\q} e_{\q}^2 \, \left\{ x  q(x, Q^2) + 
x \overline{q}(x,Q^2) \right\} ~, \hspace{4mm}
F_L(x,Q^2) = 0 ~,
\label{eq:F2partmod}
\end{equation} 
to lowest order. Such an interpretation is not valid in the limit 
$Q^2 \to 0$, where gauge invariance requires $F_2(x, Q^2) \to 0$
so that $\sigma_\mr{tot}^{\gast\p}$ remains finite. We will 
replace the DIS description by a photoproduction one in this limit. 
Hence, at small photon virtualities, the DIS process $\gast \q \ra \q$ 
should be constructed vanishingly small as compared to the contribution 
from the interaction of the hadronic component of the photon (to be 
discussed in the next section). To obtain a well-behaved DIS cross 
section in this limit, a $Q^4/(Q^2+m_\rho^2)^2$ factor is introduced. 
Here $m_\rho$ is some non-perturbative hadronic parameter, for 
simplicity identified with the $\rho$ mass. One of the $Q^2/(Q^2+m_\rho^2)$
factors is required already to give finite $\sigma_\mr{tot}^{\ga\p}$ for
conventional parton distributions, and could be viewed as a screening 
of the individual partons at small $Q^2$. The second factor is chosen to give
not only a finite but actually a vanishing $\sigma_\mr{DIS}^{\gast\p}$ 
for $Q^2 \ra 0$ in order to retain the pure photoproduction description there.
This latter factor thus is more a matter  of convenience, and other approaches
could have been pursued.
Then, in the parton model, eq.~(\ref{eq:F2}) modifies to a DIS cross section
\begin{equation}
\sigma_\mr{DIS}^{\gast\p}
\simeq\frac{4\pi^2\aem Q^2}{(Q^2+m_\rho^2)^2} \sum_{\q} e_{\q}^2 \, 
\left\{ x  q(x, Q^2) + x \overline{q}(x,Q^2) \right\}\;.
\label{eq:F2mod}
\end{equation} 
For numerical studies, the available parton distribution parameterizations 
for the proton have some lower limit of applicability in both $x$ and $Q^2$. 
For values below these minimal ones, the parton distributions are frozen at 
the lower limits.

\subsection{Photoproduction}

To first approximation, the photon is a point-like particle. However,
quantum mechanically, it may fluctuate into a (charged) 
fermion--antifermion pair. The fluctuations 
$\ga \leftrightarrow \q\qbar$ can interact strongly and therefore 
turn out to be responsible for the major part of the $\ga\p$ total 
cross section. The total rate of $\q\qbar$ fluctuations is not 
perturbatively calculable, since low-virtuality fluctuations enter a 
domain of non-perturbative QCD physics. It is therefore customary to split
the spectrum of fluctuations into a low-virtuality and a high-virtuality
part. The former part can be approximated by a sum over low-mass 
vector-meson states, customarily (but not necessarily) restricted 
to the lowest-lying vector multiplet. Phenomenologically, this 
Vector Meson Dominance (VMD) ansatz turns out to be very successful in
describing a host of data. The high-virtuality part, on the other hand, 
should be in a perturbatively calculable domain. 

In total, the photon wave function can then be written as
\begin{equation}
|\ga\rangle = c_{\mr{bare}} |\ga_{\mr{bare}}\rangle +
\sum_{V = \rho^0, \omega, \phi, \Jpsi} c_V |V\rangle +
\sum_{\q = \u, \d, \s, \c, \b} c_{\q} |\q\qbar\rangle +
\sum_{\ell = \e, \mu, \tau} c_{\ell} |\ell^+\ell^-\rangle 
\label{gammawavefunction}
\end{equation} 
(neglecting the small contribution from $\Upsilon$). In general, the 
coefficients $c_i$ depend on the scale $\mu$ used to probe the photon.
Thus $c_{\ell}^2 \approx (\aem/2\pi)(2/3) \ln(\mu^2/m_{\ell}^2)$. 
Introducing a cut-off parameter $k_0$ to separate the low- and 
high-virtuality parts of the $\q\qbar$ fluctuations, one similarly 
obtains $c_{\q}^2 \approx (\aem/2\pi) 2e_{\q}^2 \ln(\mu^2/k_0^2)$.
The VMD part corresponds to the range of $\q\qbar$ fluctuations below
$k_0$ and is thus $\mu$-independent (assuming $\mu > k_0$). 
In conventional notation $c_V^2 = 4\pi\aem/f_V^2$, with $f_V^2/4\pi$ 
determined from data to be 2.20 for $\rho^0$, 23.6 for $\omega$, 
18.4 for $\phi$ and 11.5 for $\Jpsi$ \cite{Bauer}. The $k_0$ parameter
is constrained by fits to $F_2^{\ga}$, i.e. to the parton distributions 
of the photon, to be $k_0 \simeq 0.6$~GeV \cite{SaSpdf}. (The 
fits also contain other model uncertainties, and are only logarithmically
dependent on $k_0$, so the precision is not high.) 
Finally, $c_{\mr{bare}}$ is given by
unitarity: $c_{\mr{bare}}^2 \equiv Z_3 = 1 - \sum c_V^2 -
\sum c_{\q}^2 - \sum c_{\ell}^2$. In practice, $c_{\mr{bare}}$ is
always close to unity. Usually the probing scale $\mu$ is taken to be 
the transverse momentum of a $2 \to 2$ parton-level process. 
 
\begin{figure}[t]
\begin{center}
\begin{picture}(382,100)(0,0)
  \ArrowLine(15,90)(45,70)   \ArrowLine(45,70)(85,90)
  \ArrowLine(15,10)(45,30)   \ArrowLine(45,30)(85,10)
      \Gluon(45,70)(45,30){4}{4}
     \Vertex(45,70){2}   \Vertex(45,30){2}
\DashLine(15,5)(85,5){4} \DashLine(15,95)(85,95){4}
   \GOval(10,90)(10,5)(0){0.5}  \GOval(10,10)(10,5)(0){0.5}
    \Text(1,90)[r]{$\ga$}   \Text(1,10)[r]{$\p$}
\Text(45,-9)[t]{a)} 
  \Photon(150,90)(185,70){4}{3}   \ArrowLine(230,90)(185,70)
   \Gluon(155,10)(185,30){4}{3}   \ArrowLine(185,30)(230,10)
      \Line(185,70)(185,30)
     \Vertex(185,70){2}   \Vertex(185,30){2}
\DashLine(155,5)(230,5){4}
        \GOval(150,10)(10,5)(0){0.5}
    \Text(141,90)[r]{$\ga$}   \Text(141,10)[r]{$\p$}
\Text(185,-9)[t]{b)}  
     \Photon(290,90)(320,80){4}{3}   \ArrowLine(380,90)(320,80)
  \ArrowLine(290,10)(350,30)         \ArrowLine(350,30)(380,10)
  \ArrowLine(320,80)(350,60)         \ArrowLine(350,60)(380,80)
      \Gluon(350,60)(350,30){4}{3}
     \Vertex(350,60){2}   \Vertex(350,30){2}
\DashLine(290,5)(380,5){4}
        \GOval(285,10)(10,5)(0){0.5}
    \Text(276,90)[r]{$\ga$}   \Text(276,10)[r]{$\p$}
 \Text(335,-9)[t]{c)} 
\end{picture}   
\end{center}
\vspace{2mm}
\captive%
{Contributions to hard $\ga\p$ interactions: a) VMD, 
b) direct, and c)~anomalous. Only the basic graphs are illustrated;
additional partonic activity is allowed in all three processes.
The presence of spectator jets has been indicated by dashed lines,
while full lines show partons that (may) give rise to 
high-$\pT$ jets.
\label{FigA}}
\end{figure}

The subdivision of the above photon wave function corresponds to the 
existence of three main event classes in $\ga\p$ events 
\cite{SchSjgap}, cf. Fig.~\ref{FigA}:
\begin{Enumerate}
\item The VMD processes, where the photon turns into a vector meson
before the interaction, and therefore all processes
allowed in hadronic physics may occur. This includes elastic and 
diffractive scattering as well as low-$\pT$ and high-$\pT$ 
non-diffractive events.
\item The direct processes, where a bare photon interacts with a 
parton from the proton.
\item The anomalous processes, where the photon perturbatively branches
into a $\q\qbar$ pair, and one of these (or a daughter parton thereof)
interacts with a parton from the proton. 
\end{Enumerate}
All three processes are of $\mathcal{O}(\aem)$. However, in the direct 
contribution the photon structure function is of $\mathcal{O}(1)$ and the 
hard scattering matrix elements of $\mathcal{O}(\aem)$, while the opposite 
holds for the VMD and the anomalous processes. 
The $\ell^+\ell^-$ fluctuations are not interesting for us, and 
there is thus no class associated with them.

The difference between the three classes is reflected in the beam jet 
structure. The incoming proton always gives a beam jet containing the 
partons of the proton that did not interact. On the photon side, the 
direct processes do not give a beam jet at all, since all the energy of 
the photon is involved in the hard interaction. The VMD ones (leaving 
aside the elastic and diffractive subprocesses for the moment) give a 
beam remnant just like the proton, with a moderately small `primordial
$k_{\perp}$' smearing. The anomalous processes give a beam remnant 
produced by the $\ga \to \q\qbar$ branching, with a transverse momentum 
going from $k_0$ upwards. 

Based on the different event classes discussed above, the total
photoproduction cross section can be written as
\begin{equation}
\sigma_{\mr{tot}}^{\ga\p}=
\sigma_{\mr{VMD}}^{\ga\p}+
\sigma_{\mr{direct}}^{\ga\p}+
\sigma_{\mr{anomalous}}^{\ga\p} \;.
\label{eq:evclassgp}
\end{equation}

Total hadronic cross sections show a characteristic fall-off at 
low energies and a slow rise at higher energies. This behaviour 
can be parameterized by the form 
\begin{equation}
\sigma_{\mr{tot}}^{AB}(s) = X^{AB} s^{\epsilon} + Y^{AB} s^{-\eta}
\label{sigmatotAB}
\end{equation}
for $A + B \to X$. The powers $\epsilon$ and $\eta$
are universal, with fit values \cite{DL92}
\begin{equation}
  \epsilon \approx 0.0808 ~, \qquad 
  \eta \approx 0.4525 ~,
\label{epsivalue}
\end{equation} 
while the coefficients $X^{AB}$ and $Y^{AB}$ are
process-dependent. Equation (\ref{sigmatotAB}) can be interpreted 
within Regge theory, where the first term corresponds to pomeron exchange and
gives the asymptotic rise of the cross section. 
The second term, the reggeon one, is mainly of interest at low 
energies. For the purpose of our study we do not have to rely on the Regge 
interpretation of eq.~(\ref{sigmatotAB}), but can merely consider it as 
a convenient parameterization.

The VMD part of the $\ga\p$ cross section is an obvious candidate
for a hadronic description. The diagonal VMD model suggests:
\begin{equation}
\sigma_{\mr{VMD}}^{\ga\p}(s) = 
\sum_{V=\rho^0,\omega,\phi,\Jpsi}\; 
\frac{4\pi\aem}{f_V^2}\;
\sigma_{\mr{tot}}^{V\p}(s) \;.
\label{sigmatotVMDp}
\end{equation}
Assuming an additive quark model 
the $V\p$ cross sections can be parameterized as \cite{SchSjgap}
\begin{eqnarray}
\sigma_{\mr{tot}}^{\rho^0\p}(s) 
& \approx & \sigma_{\mr{tot}}^{\omega\p}(s) 
\approx  \frac{1}{2} \left( \sigma_{\mr{tot}}^{\pi^+\p} +
  \sigma_{\mr{tot}}^{\pi^-\p} \right)
\approx 13.63 s^{\epsilon} + 31.79 s^{-\eta} ~~[\mr{mb}]  \;, \nonumber \\
\sigma_{\mr{tot}}^{\phi\p}(s) 
& \approx & 
  \sigma_{\mr{tot}}^{\K^+\p} + \sigma_{\mr{tot}}^{\K^-\p} - 
  \sigma_{\mr{tot}}^{\pi^-\p} 
  \approx 10.01 s^{\epsilon} - 1.52 s^{-\eta}~~[\mr{mb}] \;, \\
\sigma_{\mr{tot}}^{\Jpsi\p}(s) & \approx &
  \frac{m_{\phi}^2}{m_{\Jpsi}^2} \, \sigma_{\mr{tot}}^{\phi\p}(s) 
  \approx \frac{1}{10} \, \sigma_{\mr{tot}}^{\phi\p}(s)  \;, \nonumber
\label{sigmatotVp}
\end{eqnarray}
with $s$ in GeV$^2$. Adding the vector meson contributions, we arrive at
\begin{equation}
\sigma_{\mr{VMD}}^{\ga\p}(s) \approx  
53.4 s^{\epsilon} + 115 s^{-\eta}~~[\mu\mr{b}] ~.
\end{equation}

There is no compelling reason that such an ansatz should hold also for the total
$\ga\p$ cross section, but empirically a parameterization according to
\begin{equation}
\sigma_{\mr{tot}}^{\ga\p}(s) = X^{\ga\p} s^{\epsilon} + Y^{\ga\p} s^{-\eta}
= 67.7 s^{\epsilon} + 129 s^{-\eta}~~[\mu\mr{b}]
\label{eq:sigmatotgp}
\end{equation}
does a good job \cite{DL92}. For instance, these parameter values were used 
to predict the high-energy behaviour of the cross section, close to what was 
then measured by H1 and ZEUS. Thus VMD corresponds to approximately 80\% of 
the total $\ga\p$ cross section at high energies, with the remaining 20\% then 
shared among the direct and anomalous event classes.

The anomalous contribution can be written as
\begin{equation}
\sigma_{\mr{anomalous}}^{\ga\p}(s)=
\frac{\aem}{2\pi} \; \sum_\q 2 \e_\q^2 \int_{k_0^2}^{\infty} 
\frac{\d \kT^2}{\kT^2} \;
\sigma^{\q\qbar\p}(s; \kT)
\end{equation}
where the prefactor and integral over $\d\kT^2/\kT^2$ corresponds to the 
probability for the photon to split into a $\q\qbar$ state of transverse 
momenta $\pm \kT$. The cross section for this $\q\qbar$ pair to scatter 
against the proton, $\sigma^{\q\qbar\p}$, need to be modeled. 
Based only on geometrical scaling arguments (to be discussed in the 
next section), one could expect a decrease roughly like $1/\kT^2$.
This suggests an ansatz
\begin{equation}
\sigma^{\q\qbar\p}(s; \kT)=
\frac{k_{V(\q\qbar)}^2}{\kT^2} \; \sigma^{V(\q\qbar)\p}(s) ~.
\label{eq:geoscaling}
\end{equation}
The $k_{V(\q\qbar)}$ is a free parameter introduced for dimensional 
reasons. It could be associated with the typical $\kT$ inside the
vector meson $V$ formed from a $\q\qbar$ pair: $\rho^0 \approx \omega$ 
for $\u$ and $\d$, $\phi$ for $\s$, $\Jpsi$ for $\c$. As a reasonable 
ansatz, one could guess $k_{V(\q\qbar)} \approx m_V/2 \approx m_{\rho}/2$.  
(For heavier quarks, a higher mass scale is indicated, but also a 
correspondingly larger lower integration limit in $\kT^2$, so the two 
effects cancel more or less.) Fits to the total cross section at not too
high energies, with a large VMD and a small direct contributions subtracted, 
give corresponding numbers, $k_{V(\q\qbar)} \approx 0.4$~GeV for a
$k_0 \approx 0.5$~GeV. These values are here related to each other: if 
the latter were to be changed, the former would have to be retuned 
accordingly.  In the following we use this set of numbers. The anomalous 
cross section can thus be written as
\begin{equation}
\sigma_{\mr{anomalous}}^{\ga\p}(s)=
\frac{\aem}{2\pi} \; \sum_\q 2 \e_\q^2 \int_{k_0^2}^{\infty} 
\frac{\d \kT^2}{\kT^2} \;
\frac{k_{V(\q\qbar)}^2}{\kT^2} \;
\sigma^{V(\q\qbar)\p}(s)
\label{eq:anoint}
\end{equation}

To leading order, the direct events come in two kinds: QCD Compton  
$\ga \q \to \q \g$ (QCDC) and boson-gluon fusion 
$\ga \g \to \q \qbar$ (BGF). The cross sections are divergent in 
the limit $\kT \to 0$ for the outgoing parton pair. Therefore 
a lower cut-off is required, but no other specific model assumptions. 

\subsection{Combining the photoproduction processes}
\label{photop}

\begin{figure}[t]
\begin{center}
\begin{picture}(105,200)(-4,-25)
  \Photon(7,140)(45,120){4}{4}  
  \GOval(12,10)(10,5)(0){0.5}
  \DashLine(17,5)(90,5){4}
  \ArrowLine(17,13)(45,30)
  \Gluon(45,75)(45,30){4}{4}
  \ArrowLine(45,75)(45,120)
  \ArrowLine(45,30)(90,30)
  \ArrowLine(90,75)(45,75)
  \ArrowLine(45,120)(90,120) 
  \Text(0,10)[]{$\p$}
  \Text(0,140)[]{$\ga$}     
  \Text(60,50)[]{$\pT$} 
  \Text(60,95)[]{$\kT$} 
  \Text(97,30)[]{$\q'$} 
  \Text(97,75)[]{$\qbar$} 
  \Text(97,120)[]{$\q$} 
\end{picture}  
\hspace{2cm}
\begin{picture}(200,200)(0,0)
  \LongArrow(15,15)(185,15)
  \Text(195,15)[]{$\kT$} 
  \LongArrow(15,15)(15,185) 
  \Text(15,194)[]{$\pT$} 
  \SetWidth{1.5}
  \Line(50,15)(50,180)
  \Text(50,5)[]{$k_0$}
  \Line(50,50)(180,180)
  \Text(190,190)[]{$\kT = \pT$}
  \SetWidth{0.3}
  \DashLine(15,80)(50,80){2}
  \SetWidth{0.5}
  \Text(32,120)[]{VMD}
  \Text(32,110)[]{hard}   
  \Text(32,80)[]{$\pTmin$}   
  \Text(32,50)[]{VMD}
  \Text(32,40)[]{soft}
  \Text(120,55)[]{direct}
  \Text(90,140)[]{anomalous}
\end{picture}   
\end{center}
\vspace{2mm}
\captive%
{(a) Schematic graph for a hard $\ga\p$ process, illustrating
the concept of two different scales. 
(b) The allowed phase space for this process, with one subdivision
into event classes.
\label{FigB}}
\end{figure}

The VMD, direct and anomalous classes have so far been considered 
separately. The complete physics picture presumably would provide 
smooth transitions between the various possibilities. To understand 
the relation between the processes, consider the simple graph of 
Fig.~\ref{FigB}a. There two transverse momentum scales, $\kT$ and 
$\pT$, are introduced. 
Here $\kT$ is related to the $\ga \to \q\qbar$ vertex while 
$\pT$ is the hardest QCD $2 \to 2$ subprocess of the ladder between 
the photon and the proton. (Further softer partons in the ladders are 
omitted for clarity.) The allowed phase space can then conveniently be 
represented by a two-dimensional plane, Fig.~\ref{FigB}b. The region 
$\kT < k_0$ corresponds to a small transverse momentum at the 
$\ga \to \q\qbar$ vertex, and thus to VMD processes. For $\kT > k_0$, 
the events are split along the diagonal $\kT = \pT$. If $\kT > \pT$, 
the hard $2 \to 2$ process of Fig.~\ref{FigB}a is $\ga\g \to \q\qbar$, 
and the lower part of the graph is part of the leading log QCD evolution 
of the gluon distribution inside the proton. These events are direct ones. 
If $\pT > \kT$, on the other hand, the hard process is 
$\qbar \q' \to \qbar \q'$, and the $\ga \to \q\qbar$ vertex builds 
up the quark distribution inside a photon. These events are thus 
anomalous ones.

It should be remembered that the classification is only simple 
away from the border regions. When $\kT \approx \pT$, say, 
a description either in terms of anomalous or direct interactions
would be possible. Also higher-order corrections will blur the
picture, although pragmatic separations normally can be found.

A comment on the choice of variables. Instead of the squared transverse
momenta $\kT^2$ and $\pT^2$ one could equally well have imagined
the squared virtualities $-k^2$ and $-p^2$. However, in many processes
several Feynman graphs contribute, both in the $t$-, $u$- and 
$s$-channel. The association of $p^2 = \hat{t}$ or $p^2 = \hat{u}$
is then not unique, while $\pT^2 = \hat{t} \hat{u} / \hat{s}$
is well-defined and coincides with $-\hat{t}$ or $-\hat{u}$ in the
respective singular limit. Whether events are classified by one kind
of scale or another should not be relevant, at least so long as one
is consistent and stays with a leading-order description. Only for the 
virtual photon do we use the scale $Q^2 = -q^2 = -\hat{t}$ rather than 
$q_{\perp}^2$, in order to stay in line with conventional notation, and 
because there is no scale choice ambiguity here. 

What complicates the picture is that an event may contain several
interactions, once one considers an incoming particle
as a composite object with several partons that may interact,
more or less independently of each other, with partons from the
other incoming particle. Such a multiple parton--parton interaction 
scenario is familiar already from $\p\p$ physics \cite{multint}. 
Here the jet cross section, above some $\pTmin$ scale of the order 
of 2~GeV, increases faster with energy than the total cross section.
Above an energy of a few hundred GeV the calculated jet cross section 
is larger than the observed total one. Multiple interactions offers a 
solution to this apparent paradox, by squeezing a larger number of 
jet pairs into the average event, a process called unitarization 
or eikonalization. The perturbative jet cross section
can then be preserved, at least down to $\pTmin$, but in
the reinterpreted inclusive sense. At the same time, the unitarization
plays a crucial r\^ole in taming the growth of the total 
cross section. 

The composite nature of hadrons also fills another function: it
regularizes the singularity of perturbative cross sections, such as 
$\q\g \to \q\g$, in the limit $\pT \to 0$. Perturbative calculations 
assume free colour charges in the initial and final states of the 
process, while the confinement in hadrons introduces some typical colour
neutralization distance. It is the inverse of this scale that appears
as some effective cutoff scale $\pTmin \simeq 2$~GeV, most likely
with a slow energy dependence \cite{Johann}. One possible 
parameterization is
\begin{equation}
\pTmin(s) = (1.9~{\mr{GeV}}) \left(
\frac{s}{1~\mr{TeV}^2} \right)^{\epsilon} ~,
\label{eq:pTminsdep}
\end{equation}
with the same $\epsilon$ as in eq.~(\ref{epsivalue}), since
the rise of the total cross section with energy via Regge 
theory is related to the small-$x$ behaviour of parton distributions
and thus to the density of partons.

Now, if an event contains interactions at several different $\pT$
scales, standard practice is to classify this event by its hardest
interaction. Several reasons can be given; one is that a softer 
interaction may be confused with QCD radiation emitted from the
harder one, and thus cannot be identified on an exclusive basis.
With this prescription, the cross section for an event of scale $\pT$ 
is the naive jet cross section at this $\pT$ scale \textit{times} 
the probability that the event contained no interaction at a scale
above $\pT$. The latter defines a form factor, related to
probability conservation, analogously to but not equivalent with
the Sudakov form factor of parton showers. At large $\pT$ values
the probability of having an even larger $\pT$ is small, i.e.
the form factor is close to unity, and the perturbative cross 
section is directly preserved in the event rate. At lower
$\pT$ values, the likelihood of a larger $\pT$ is increased,
i.e. the form factor becomes smaller than unity, and the rate of 
events classified by this $\pT$ scale falls below the perturbative
answer.  

We expect this picture to hold also for the VMD part of the photon,
since this is clearly in the domain of hadronic physics. Thus,
in the VMD domain $\kT < k_0$, the region of large $\pT$ in 
Fig.~\ref{FigB}b is populated according to perturbation theory,
though with nonperturbative input to the parton distributions.
The region of smaller $\pT$ is suppressed, since the form 
factor here drops significantly below unity. 

As one moves away from the `pure' VMD states, such as the $\rho^0$,
much of the same picture could well hold. Interactions at a
larger $\kT$ value could be described in terms of some $\rho'$ state.
The uncertainty relation gives us that a state of virtuality 
$\simeq \kT$ has a maximal size $\simeq 1 / \kT$ and thus spans an 
area $\propto 1 / \kT^2$. In a naive picture, 
the $\rho' \p$ cross section would then drop with increasing
$\kT$, but flatten out at a value given by the proton size. 
We know that this is not the scaling observed in nature, however,
where e.g. the $\Jpsi \p$ cross section is much below the 
$\phi \p$ one \cite{SaSgaga}. Rather it appears that the cross
section is proportional to the area of the state interacting
with the proton, i.e. a (kind of) geometrical scaling. 
Such a behaviour becomes understandable (but not easily predicted
in detail) when one remembers that
the colour neutralization distance inside a more virtual photon
state is also reduced, so that the interactions in general tend to 
be weakened by interference effects not included in the simple 
perturbative cross sections. This could then be the origin for
a geometrical scaling like the one in eq.~(\ref{eq:geoscaling}). 

Alternatively to the geometric approach, the anomalous and direct 
cross sections for such a $\kT$ state could be calculated. If these 
latter are larger than the geometrical scaling answer, one could 
expect multiple interaction effects to become important and couple 
the two event classes. That is, the presence of an anomalous 
interaction of some large $\pT$ would preempt the event from being 
classifiable as a direct one with a lower $\pT$, in analogy with 
the discussion on hadronic physics above. One can even attempt 
eikonalized descriptions \cite{Vietri}, although such require a 
number of assumptions to be made.

Calculating the perturbative anomalous cross section in the region
given by $\pT > \max(\kT, \pTmin(s))$, the geometric scaling answer 
is exceeded
for some region $\kT \lessim k_1$, with $k_1 \approx 2-4$~GeV
(higher for higher energies). Only for $\kT > k_1$ is the jet cross 
section dropping below the geometric scaling one. At these larger 
$\kT$ values, the direct rate dominates over the anomalous. As a 
convenient but rather arbitrary choice, for subsequent studies we put 
$k_1 = \pTmin(s)$, with the latter given by eq.~(\ref{eq:pTminsdep}). 
This value lies below the crossover point noted above, but the   
simple modified geometrical scaling picture is not precise enough 
that it can be trusted too far, so a separate $k_1$ parameterization
would seem excessive. 

\begin{figure}[t]
\begin{center}
\begin{picture}(200,200)(0,0)
  \LongArrow(15,15)(185,15)
  \Text(195,15)[]{$\kT$} 
  \LongArrow(15,15)(15,185) 
  \Text(15,194)[]{$\pT$} 
  \SetWidth{1.5}
  \Line(50,15)(50,180)
  \Text(50,5)[]{$k_0$}
  \Line(100,15)(100,100)
  \Text(100,5)[]{$k_1$}
  \Line(100,100)(180,180)
  \Text(190,190)[]{$\kT = \pT$}
  \Text(32,90)[]{VMD}
  \Text(75,100)[]{GVMD}
  \Text(140,60)[]{direct}
\end{picture}   
\end{center}
\vspace{2mm}
\captive%
{An alternative classification of the phase space in Fig.~\ref{FigB},
which better takes into account unitarization effects. 
\label{FigC}}
\end{figure}

The final scenario is illustrated in Fig.~\ref{FigC}.
The bulk of the cross section, in the region $\kT < k_1$,
is now described by the photon interacting as dense, hadronic states,
VMD for $\kT \lessim k_0$ and Generalized VMD (GVMD) for
$k_0 \lessim \kT \lessim k_1$. The total VMD cross section is 
given by the pomeron-type ansatz, while the jet cross section
can be obtained from the parton distributions of the respective
vector meson state. Correspondingly, the GVMD states have a
total cross section based on Pomeron considerations and a jet
cross section now based on the anomalous part of the parton
distributions of the photon. In principle, an eikonalization
should be performed for each GVMD state separately, but in 
practice that would be overkill. Instead the whole region is
represented by one single state per quark flavour, with a jet 
production given by the full anomalous part of the photon 
distributions. By extending this jet rate also to the region 
$\kT > k_1$, one does introduce a small mismatch between the 
low-$\pT$ and high-$\pT$ descriptions, but retains the correct 
full jet rate at high $\pT$'s. Implicitly, the presence of a 
spectrum of $\kT$ states can be taken into account by the choice 
of a realistic $\pT$ spectrum of the photon beam remnant, maybe 
still without the full correlation structure between the high-$\pT$ 
and low-$\pT$ parts of the individual events.

Thus, post facto, the approximate validity of a Regge theory ansatz 
for $\sigma_{\mr{tot}}^{\ga\p}$ is making sense. Above $k_1$
only the direct cross section need be considered, since here the
anomalous one is negligibly small, at least in terms of total cross 
sections. (As noted above, we have actually chosen to lump it with
the other GVMD contributions, so as not to lose the jet rate itself.) 
Numerically, the recipe of extending the anomalous contribution to 
infinity according to a GVMD scaling recipe, as is done in 
eq.~(\ref{eq:anoint}), is about equally good. The latter may involve 
some double-counting with the direct cross section, but not more than 
falls within the general uncertainty of the geometric scaling and 
eikonalization game. 

\subsection{DIS revisited}

\begin{figure}[t]
\begin{center}
\begin{picture}(105,220)(-4,-25)
  \ArrowLine(5,165)(45,165)
  \ArrowLine(45,165)(90,180)
  \Photon(45,165)(45,120){4}{4}  
  \GOval(12,10)(10,5)(0){0.5}
  \DashLine(17,5)(90,5){4}
  \ArrowLine(17,13)(45,30)
  \Gluon(45,75)(45,30){4}{4}
  \ArrowLine(45,75)(45,120)
  \ArrowLine(45,30)(90,30)
  \ArrowLine(90,75)(45,75)
  \ArrowLine(45,120)(90,120) 
  \Text(0,10)[]{$\p$}
  \Text(0,165)[]{$\e$}
  \Text(30,140)[]{$\ga^*$}     
  \Text(30,50)[]{$\g$}     
  \Text(60,50)[]{$\pT$} 
  \Text(60,95)[]{$\kT$} 
  \Text(60,140)[]{$Q$} 
  \Text(97,30)[]{$\q'$} 
  \Text(97,75)[]{$\qbar$} 
  \Text(97,120)[]{$\q$} 
  \Text(97,180)[]{$\e'$} 
\end{picture} 
\hspace{2cm} 
\begin{picture}(200,200)(0,0)
  \LongArrow(15,15)(185,15)
  \Text(195,15)[]{$\kT$} 
  \LongArrow(15,15)(15,185) 
  \Text(15,194)[]{$\pT$} 
  \SetWidth{1.5}
  \Line(15,15)(180,180)
  \Line(100,15)(100,180)
  \Text(100,5)[]{$Q$}
  \Text(190,190)[]{$\kT = \pT$}
  \Text(70,35)[]{LO DIS}
  \Text(150,60)[]{QCDC+BGF}
  \Text(55,100)[]{non-DGLAP}
  \Text(135,170)[]{photoprod}
\end{picture}   
\end{center}
\vspace{2mm}
\captive%
{(a) Schematic graph for a hard $\ga^*\p$ process, illustrating
the concept of three different scales. (b) Event classification
in the large-$Q^2$ limit.
\label{fig:classDIS}}
\end{figure}

In DIS, the photon virtuality $Q^2$ 
introduces a further scale to the process, i.e. one goes from
Fig.~\ref{FigB}a to Fig.~\ref{fig:classDIS}a.
The traditional DIS region is the strongly ordered one,
$Q^2 \gg \kT^2 \gg \pT^2$, where DGLAP-style evolution \cite{DGLAP}
is responsible for the event structure. As above, ideology 
wants strong ordering, while real life normally is
based on ordinary ordering $Q^2 > \kT^2 > \pT^2$.
Then the parton-model description 
of $F_2(x,Q^2)$ in eq.~(\ref{eq:F2partmod}) is a very good first 
approximation. The problems come when the ordering is no longer 
well-defined, i.e. either when the process contains several large 
scales or when $Q^2 \to 0$. In these regions, an $F_2(x,Q^2)$ may
still be defined by eq.~(\ref{eq:F2}), but its physics interpretation 
is not obvious.
 
Let us first consider a large $Q^2$, where a possible classification
is illustrated in Fig.~\ref{fig:classDIS}b. The regions $Q^2 > \pT^2 > \kT^2$ 
and $\pT^2 > Q^2 > \kT^2$ correspond to non-ordered emissions, that 
then go beyond DGLAP validity and instead have to be described by the 
BFKL \cite{BFKL} or CCFM \cite{CCFM} equations, see e.g. \cite{LDC}. 
Normally one expects such cross sections to be small at large $Q^2$. 
The (sparsely populated) region $\pT^2 > \kT^2 > Q^2$ can be viewed 
as the interactions of a resolved (anomalous) photon.

The region $\kT^2 > Q^2 \gg 0$ and $\kT^2 > \pT^2$ contains the 
${\mathcal O}(\as)$ corrections to the lowest-order (LO) DIS process 
$\ga^* \q \to \q$, namely QCD Compton $\ga^* \q \to \q \g$ 
and boson-gluon fusion $\ga^* \g \to \q \qbar$. These 
are nothing but the direct processes  $\ga \q \to \q \g$ and 
$\ga \g \to \q \qbar$ extended to virtual photons. The borderline
$\kT^2 > Q^2$ is here arbitrary --- also processes with $\kT^2 < Q^2$
could be described in this language. In the parton model, this whole 
class of events are implicitly included in $F_2$, and are related
to the logarithmic scaling violations of the parton distributions.
The main advantage of a separation at $\kT = Q$ thus comes from the 
matching to photoproduction. Also the exclusive modeling of events, 
with the attaching of parton showers of scale $Q^2$ to DIS events,
is then fairly natural.

The DIS cross section thus is subdivided into
\begin{equation}
\sigma_\mr{tot}^{\gast\p}
 \simeq  
\frac{4\pi^2\aem Q^2}{(Q^2+m_\rho^2)^2} F_2(x,Q^2) = \sigma_{F_2}^{\gast\p}
 \simeq  \sigma_{\mr{DIS}}^{\gast\p} =
\sigma_{\mr{LO\,DIS}}^{\gast\p} +
\sigma_{\mr{QCDC}}^{\gast\p} + \sigma_{\mr{BGF}}^{\gast\p}
\;.
\label{eq:DIScomp}
\end{equation}
The $\sigma_{\mr{DIS}}^{\gast\p}$ is given by eq.~(\ref{eq:F2mod}),
while the last two terms are well-defined by an integration of the 
respective matrix element \cite{lutp9911}. When extended to small $Q^2$, 
these two terms will increase in importance, and one may eventually 
encounter a $\sigma_{\mr{LO\,DIS}}^{\gast\p} < 0$, if calculated
by a subtraction of the QCDC and BGF terms from the total DIS cross 
section. However, here we expect the correct answer not to be a 
negative number but an exponentially suppressed one, by a Sudakov form 
factor. This modifies the cross section: 
\begin{equation}
\sigma_{\mr{LO\,DIS}}^{\gast\p} = \sigma_{\mr{DIS}}^{\gast\p} -
\sigma_{\mr{QCDC}}^{\gast\p} - \sigma_{\mr{BGF}}^{\gast\p}
~~ \longrightarrow ~~ 
\sigma_{\mr{DIS}}^{\gast\p} \; \exp \left( - \frac{%
\sigma_{\mr{QCDC}}^{\gast\p} + \sigma_{\mr{BGF}}^{\gast\p}}%
{\sigma_{\mr{DIS}}^{\gast\p}} \right) \;.
\label{eq:LODISmod}
\end{equation}
Since we here are in a region where
$\sigma_{\mr{DIS}}^{\gast\p} \ll \sigma_{F_2}^{\gast\p}$,
i.e. where the DIS cross section is no longer the dominant one, this
change of the total DIS cross section is not essential. Even more,
for $Q^2 \to 0$ we know that the direct processes should survive whereas 
the lowest-order DIS one has to vanish. Since eq.~(\ref{eq:F2mod})
ensures that $\sigma_{\mr{DIS}}^{\gast\p} \to 0$  in this limit,
it also follows that $\sigma_{\mr{LO\,DIS}}^{\gast\p}$ does so.   

\subsection{From Real to Virtual Photons}

It is now time to try to combine the different aspects of the photon, 
to provide an answer that smoothly interpolates between the 
photoproduction and DIS descriptions, in a physically sensible way.

A virtual photon has a reduced probability to fluctuate into a vector 
meson state, and this state has a reduced interaction probability. 
This can be modeled with the traditional dipole factors 
\cite{dipolevirt} 
\begin{equation}
\sigma_{\mr{VMD}}^{\gast\p}(s,Q^2) = 
\sum_{V=\rho^0,\omega,\phi,\Jpsi}\; 
\frac{4\pi\aem}{f_V^2}\;
\left( \frac{m_V^2}{m_V^2 + Q^2} \right)^2 \; 
\sigma_{\mr{tot}}^{V\p}(s) \;.
\label{sigmatotVMDpvirt}
\end{equation}
Similarly, the GVMD states are affected,
\begin{equation}
\sigma_{\mr{GVMD}}^{\gast\p}(s,Q^2)=
\frac{\aem}{2\pi} \; \sum_\q 2 \e_\q^2 \int_{k_0^2}^{k_1^2} 
\frac{\d \kT^2}{\kT^2} \;
\left( \frac{4 \kT^2}{4 \kT^2+Q^2} \right)^2 \;
\frac{k_{V(\q\qbar)}^2}{\kT^2} \;
\sigma^{V(\q\qbar)\p}(s) \;,
\end{equation}
where a relation $2 \kT \simeq m$ is assumed.

The above generalization to virtual photons does not address the issue 
of longitudinal photons. Their interactions vanish in the limit 
$Q^2 \to 0$, but can well give a non-negligible contribution at
finite $Q^2$ \cite{longitdata}. A common approach is to attribute the 
longitudinal cross section with an extra factor of $r_V=a_V Q^2/m_V^2$ 
relative to the transverse one \cite{longitmod}, where $a_V$ is some 
unknown parameter to be determined from data. Such an ansatz only 
appears reasonable for moderately small $Q^2$, however, so following 
the lines of our previous study of jet production by virtual 
photons~\cite{lutp9911}, we will try the two alternatives 
\begin{eqnarray}
r_1(m_V^2, Q^2) & = & a \frac{4 m_V^2 Q^2}{(m_V^2 + Q^2)^2}~,\label{eq:rV1}\\
r_2(m_V^2, Q^2) & = & a \frac{4 Q^2}{(m_V^2 + Q^2)}~.\label{eq:rV2}
\end{eqnarray}
While $r_1$ vanishes for high $Q^2$, $r_2$ approaches the constant 
value $a$. The above VMD expressions are again extended to GVMD by the 
identification $m_V \approx 2\kT$. The cross sections can then be 
written as
\begin{eqnarray}
\sigma_{\mr{VMD}}^{\gast\p}(W^2, Q^2) & = &
\sum_{V=\rho^0,\omega,\phi,\Jpsi}\; 
\frac{4\pi\aem}{f_V^2}\;
\left[ 1+r_i(m_V^2, Q^2) \right] 
\; \times \nonumber \\
& \times &
\left( \frac{m_V^2}{m_V^2 + Q^2} \right)^2 \; 
\sigma_{\mr{tot}}^{V\p}(W^2) \;,
\label{sigmatotVMDpTL} \\
\sigma_{\mr{GVMD}}^{\gast\p}(W^2, Q^2) & = &
\frac{\aem}{2\pi} \; \sum_\q 2 \e_\q^2 \int_{k_0^2}^{k_1^2} 
\frac{\d \kT^2}{\kT^2} \;
\left[ 1+r_i(4 \kT^2, Q^2)\right]
\; \times \nonumber \\
& \times &
\left( \frac{4 \kT^2}{4 \kT^2+Q^2} \right)^2 
\frac{k_{V(\q\qbar)}^2}{\kT^2} \;
\sigma^{V(\q\qbar)\p}(W^2) \;.
\label{eq:sigmatotAnopTL}
\end{eqnarray}
Note that we also here replaced $s$ by the conventional DIS variable
$W^2$; obviously both refer to the same $\gast\p$ squared invariant
mass.

The extrapolation to $Q^2 > 0$ is trivial for the direct processes,
which coincide with the DIS QCDC and BGF processes. The matrix elements
contain all the required $Q^2$ dependence, with a smooth behaviour in
the $Q^2 \to 0$ limit. They are to be applied to the region
$\kT > \max(k_1, Q)$ (and $\kT > \pT$, as usual). 

Remains the LO DIS process. It is here that one could encounter an
overlap and thereby double-counting with the VMD and GVMD processes.
Comparing Fig.~\ref{fig:classDIS}b with Fig.~\ref{FigC}, one may note that
the region $\pT > \kT$ involves no problems, since we have made no
attempt at a non-DGLAP DIS description but cover this region entirely
by the VMD/GVMD descriptions. Also, if $Q > k_1$, then the region 
$k_1 < \kT < Q$ (and $\kT > \pT$) is covered by the DIS process only.
So it is in the corner $\kT < k_1$ that the overlap can occur. 
If $Q^2$ is very small, the exponential factor in 
eq.~(\ref{eq:LODISmod}) makes the DIS contribution too small to worry 
about. Correspondingly, if $Q^2$ is very big, the VMD/GVMD contributions
are too small to worry about. Furthermore, a large $Q^2$ implies
a Sudakov factor suppression of a small $\kT$ in the DIS description. 
If $W^2$ is large, the multiple-interaction discussions above are 
relevant for the VMD/GVMD states: the likelihood of an interaction at 
large $\pT$ will preempt the population of the low-$\pT$ region. 

In summary, it is only in the region of intermediate $Q^2$ and rather
small $W^2$ that we have reason to worry about a significant
double-counting. As we shall see, it is indeed
in this region that the VMD/GVMD contribution does not appear to dampen
quite as fast as data indicates. Typically, this is the region where
$x \approx Q^2/(Q^2 + W^2)$ is not close to zero, and where $F_2$ is
dominated by the valence-quark contribution. The latter behaves roughly
$\propto (1-x)^n$, with an $n$ of the order of 3 or 4. Therefore
we will introduce a corresponding damping factor to the VMD/GVMD terms.
The real damping might be somewhat different but, since small $W$ values
are not our prime interest, we rest content with this approximate form.

In total, we have now arrived at our ansatz for all $Q^2$
\begin{eqnarray}
\sigma_\mr{tot}^{\gast\p} & = &
\sigma_{\mr{DIS}}^{\gast\p} \; \exp \left( - \frac{%
\sigma_{\mr{QCDC}}^{\gast\p} + \sigma_{\mr{BGF}}^{\gast\p}}%
{\sigma_{\mr{DIS}}^{\gast\p}} \right) +
\sigma_{\mr{QCDC}}^{\gast\p} + \sigma_{\mr{BGF}}^{\gast\p} 
\nonumber \\
& + &\left( \frac{W^2}{Q^2 + W^2} \right)^n \left(
\sigma_{\mr{VMD}}^{\gast\p} + 
\sigma_{\mr{GVMD}}^{\gast\p} \right) \;,
\label{eq:siggastptot}
\end{eqnarray}
where the DIS, VMD and GVMD terms are given by eqs.~(\ref{eq:F2mod}),
(\ref{sigmatotVMDpTL}) and (\ref{eq:sigmatotAnopTL}), respectively,
and the QCDC and BGF terms by direct integration of the respective
matrix elements for the region $\kT > \max(k_1, Q)$. To keep the
terminology reasonably compact, also for the $\gast\gast$ case below,
we will use res as shorthand for the resolved VMD plus GVMD 
contributions and dir as shorthand for the QCDC and BGF processes.
Then eq.~(\ref{eq:siggastptot}) simplifies to
\begin{equation}
\sigma_\mr{tot}^{\gast\p}  = 
\sigma_{\mr{DIS}}^{\gast\p} \; \exp \left( - \frac{%
\sigma_{\mr{dir}}^{\gast\p}}{\sigma_{\mr{DIS}}^{\gast\p}} \right) +
\sigma_{\mr{dir}}^{\gast\p} +
\left( \frac{W^2}{Q^2 + W^2} \right)^n 
\sigma_{\mr{res}}^{\gast\p} \;.
\label{eq:siggastptotsimple}
\end{equation}

\subsection{Photon--photon collisions}
\label{sec:5}

\begin{figure}[t]
\begin{center}
\begin{picture}(382,100)(0,0)
  \ArrowLine(15,90)(45,70)   \ArrowLine(45,70)(85,90)
  \ArrowLine(15,10)(45,30)   \ArrowLine(45,30)(85,10)
      \Gluon(45,70)(45,30){4}{4}
     \Vertex(45,70){2}   \Vertex(45,30){2}
\DashLine(15,5)(85,5){4} \DashLine(15,95)(85,95){4}
   \GOval(10,90)(10,5)(0){0.5}  \GOval(10,10)(10,5)(0){0.5}
    \Text(1,90)[r]{$\ga$}   \Text(1,10)[r]{$\ga$}
\Text(45,-9)[t]{a)} 
  \Photon(150,90)(185,70){4}{3}   \ArrowLine(230,90)(185,70)
   \Gluon(155,10)(185,30){4}{3}   \ArrowLine(185,30)(230,10)
      \Line(185,70)(185,30)
     \Vertex(185,70){2}   \Vertex(185,30){2}
\DashLine(155,5)(230,5){4}
        \GOval(150,10)(10,5)(0){0.5}
    \Text(141,90)[r]{$\ga$}   \Text(141,10)[r]{$\ga$}
\Text(185,-9)[t]{b)}  
     \Photon(290,90)(320,80){4}{3}   \ArrowLine(380,90)(320,80)
  \ArrowLine(290,10)(350,30)         \ArrowLine(350,30)(380,10)
  \ArrowLine(320,80)(350,60)         \ArrowLine(350,60)(380,80)
      \Gluon(350,60)(350,30){4}{3}
     \Vertex(350,60){2}   \Vertex(350,30){2}
\DashLine(290,5)(380,5){4}
        \GOval(285,10)(10,5)(0){0.5}
    \Text(276,90)[r]{$\ga$}   \Text(276,10)[r]{$\ga$}
 \Text(335,-9)[t]{c)} 
\end{picture} \\
\vspace{1cm}   
\begin{picture}(382,100)(0,0)
  \Photon(10,90)(45,70){4}{3}   \ArrowLine(85,90)(45,70)
  \Photon(10,10)(45,30){4}{3}   \ArrowLine(45,30)(85,10)
      \Line(45,70)(45,30)
     \Vertex(45,70){2}   \Vertex(45,30){2}
    \Text(1,90)[r]{$\ga$}   \Text(1,10)[r]{$\ga$}
\Text(45,-9)[t]{d)} 
     \Photon(150,90)(200,75){4}{5}   \ArrowLine(200,75)(230,90)
     \Photon(150,10)(180,20){4}{3}   \ArrowLine(230,10)(180,20)
  \ArrowLine(180,20)(200,35)         \Gluon(200,35)(230,20){4}{3}    
      \Line(200,75)(200,35)
     \Vertex(200,75){2}   \Vertex(200,35){2}
    \Text(141,90)[r]{$\ga$}   \Text(141,10)[r]{$\ga$}
\Text(185,-9)[t]{e)}  
     \Photon(290,90)(320,80){4}{3}   \ArrowLine(380,90)(320,80)
  \ArrowLine(320,80)(350,65)         \ArrowLine(350,65)(380,80) 
     \Photon(290,10)(320,20){4}{3}   \ArrowLine(380,10)(320,20)
  \ArrowLine(320,20)(350,35)         \ArrowLine(350,35)(380,20)    
      \Gluon(350,65)(350,35){4}{3}
     \Vertex(350,65){2}   \Vertex(350,35){2}
    \Text(276,90)[r]{$\ga$}   \Text(276,10)[r]{$\ga$}
\Text(335,-9)[t]{f)} 
\end{picture}
\end{center}
\vspace{2mm}   
\captive%
{Contributions to hard real $\ga\ga$ interactions: a) VMD$\times$VMD, 
b)~VMD$\times$direct, c) VMD$\times$anomalous, 
d) direct$\times$direct, e) direct$\times$anomalous, 
and f)~anomalous$\times$anomalous. 
Notation as in Fig.~\protect\ref{FigA}.
\label{FigE}}
\end{figure}

A generalization of the $\ga\p$ picture to $\ga\ga$ events is 
obtained by noting that each of the two incoming photons is
described by a wave function of the type given in 
eq.~(\ref{gammawavefunction}). In total, there are therefore 
three times three
event classes. By symmetry, the `off-diagonal' combinations appear
pairwise, so for real photons the number of distinct classes is only 
six. These are, cf. Fig.~\ref{FigE}:
\begin{Enumerate}
\item VMD$\times$VMD: both photons turn into hadrons, and the processes
are therefore the same as allowed in hadron--hadron collisions.
\item VMD$\times$direct: a bare photon interacts with the partons of the
VMD photon.
\item VMD$\times$anomalous: the anomalous photon perturbatively 
branches into a $\q\qbar$ pair, and one of these (or a daughter parton 
thereof) interacts with a parton from the VMD photon.
\item Direct$\times$direct: the two photons directly give a quark pair,
$\ga\ga \to \q\qbar$. Also lepton pair production is allowed,
$\ga\ga \to \ell^+\ell^-$, but will not be considered by us.
\item Direct$\times$anomalous: the anomalous photon perturbatively 
branches into a $\q\qbar$ pair, and one of these (or a daughter parton 
thereof) directly interacts with the other photon. 
\item Anomalous$\times$anomalous: both photons perturbatively branch 
into $\q\qbar$ pairs, and subsequently one parton from each photon 
undergoes a hard interaction.
\end{Enumerate}
Most of the above classes above are pretty much the same as  allowed in 
$\ga\p$ events, since the interactions of a VMD or anomalous photon and 
those of a proton are about the same. Only the direct$\times$direct
class offer a new hard subprocess.

\begin{figure}[t]
\begin{center}
\begin{minipage}[b]{5cm}
\begin{picture}(114,140)(-12,-25)
  \Photon(0,120)(45,120){4}{4}  
  \ArrowLine(90,75)(45,75)
  \ArrowLine(45,75)(45,120)
  \ArrowLine(45,120)(90,120) 
  \Gluon(45,75)(45,30){4}{4}
  \ArrowLine(90,-15)(45,-15)
  \ArrowLine(45,-15)(45,30)
  \ArrowLine(45,30)(90,30) 
  \Photon(0,-15)(45,-15){4}{4}  
  \Text(-7,120)[]{$\ga$}
  \Text(-7,-15)[]{$\ga$}     
  \Text(60,5)[]{$\kTtwo$} 
  \Text(60,50)[]{$\pT$} 
  \Text(60,95)[]{$\kTone$}
  \Text(97,-15)[]{$\qbar'$} 
  \Text(97,30)[]{$\q'$} 
  \Text(97,75)[]{$\qbar$} 
  \Text(97,120)[]{$\q$} 
\end{picture}  
\end{minipage}
\hspace{0.5cm}
\begin{minipage}[b]{8.5cm}
1. VMD$\times$VMD: $\kTone, \kTtwo < k_0$, arbitrary $\pT$\\
2. VMD$\times$direct: $\kTone < k_0 < \pT < \kTtwo$\\ 
 + ($1 \leftrightarrow 2$)\\
3. VMD$\times$anomalous: $\kTone < k_0 < \kTtwo < \pT$\\
 + ($1 \leftrightarrow 2$)\\
4. Direct$\times$direct: $k_0 < \kTone = \kTtwo$\\
5. Direct$\times$anomalous: $k_0 < \kTone < \pT < \kTtwo$\\
 + ($1 \leftrightarrow 2$)\\
6. Anomalous$\times$anomalous: $k_0 < \kTone, \kTtwo < \pT$\\
\hspace*{10mm}
\end{minipage}
\end{center}
\vspace{2mm}
\captive%
{Schematic graph for a hard $\ga\ga$ process, showing the three 
different scales. To the right is shown the relation to the six 
classes in the text.
\label{FigF}}
\end{figure}

The main parton-level processes that occur in the above classes are:
\begin{Itemize}
\item The `direct' processes $\ga\ga \to \q\qbar$ only occur 
in class 4.
\item The `single-resolved' processes $\ga\q \to \q\g$ and 
$\ga\g \to \q\qbar$ occur in classes 2 and 5.
\item The `double-resolved' processes $\q\q' \to \q\q'$ (where $\q'$ 
may also represent an antiquark), $\q\qbar \to \q'\qbar'$,
$\q\qbar \to \g\g$, $\q\g \to \q\g$, $\g\g \to \q\qbar$ and
$\g\g \to \g\g$ occur in classes 1, 3 and 6.
\end{Itemize}
The classification of a generic Feynman diagram by the different possible
components is illustrated in Fig.~\ref{FigF}. The appearance of more 
scales makes it infeasible to draw diagrams like Fig.~\ref{FigB}b,
but the principles are the same. 

Also the extension to virtual photons follows from the
$\gast\p$ formalism above, but now with (up to) five scales to keep 
track of: $\pT$, $\kTone$, $\kTtwo$, $Q_1$ and $Q_2$. First consider
the three by three classes present already for real photons, which
remain nine distinct ones for $Q_1^2 \neq Q_2^2$. Each VMD or GVMD
state is associated with its dipole damping factor and its correction
factor for the longitudinal contribution. The QCDC and BGF matrix
elements involving one direct photon on a VMD or a GVMD state explicitly
contain the dependence on the direct photon virtuality, separately
given for the transverse and the longitudinal contributions. Also the
direct$\times$direct matrix elements are known for the four possible
transverse/longitudinal combinations. Some examples should be enough:
\begin{eqnarray}
\sigma_{\mr{VMD}\times\mr{GVMD}}^{\gast\gast}(W^2, Q_1^2, Q_2^2) 
& = &
\sum_{V_1=\rho^0,\omega,\phi,\Jpsi} \; \frac{4\pi\aem}{f_{V_1}^2} \;
\left[ 1+r_i(m_{V_1}^2, Q_1^2) \right] \; 
\left( \frac{m_{V_1}^2}{m_{V_1}^2 + Q_1^2} \right)^2 \; \times 
\nonumber \\
& \times &
\frac{\aem}{2\pi} \; \sum_\q 2 \e_\q^2 \int_{k_0^2}^{k_1^2} 
\frac{\d \kT^2}{\kT^2} \; \left[ 1+r_i(4 \kT^2, Q_2^2)\right] \; \times
\nonumber \\
& \times &
\left( \frac{4 \kT^2}{4 \kT^2+Q_2^2} \right)^2 
\frac{k_{V_2(\q\qbar)}^2}{\kT^2} \;
\sigma^{V_1 V_2(\q\qbar)}(W^2) \; ,
\label{sigmatotVMDGVMD} \\
\sigma_{\mr{VMD}\times\mr{dir}}^{\gast\gast}(W^2, Q_1^2, Q_2^2) 
& = &
\sum_{V_1=\rho^0,\omega,\phi,\Jpsi} \; \frac{4\pi\aem}{f_{V_1}^2} \;
\left[ 1+r_i(m_{V_1}^2, Q_1^2) \right] \; 
\left( \frac{m_{V_1}^2}{m_{V_1}^2 + Q_1^2} \right)^2 \; \times 
\nonumber \\
& \times &
\left( \sigma_{\mr{T}}^{V_1\times\mr{dir}}(W^2, Q_2^2) +
\sigma_{\mr{L}}^{V_1\times\mr{dir}}(W^2, Q_2^2) \right) \;.
\label{eq:sigmatotVMDdir}
\end{eqnarray}

To this should be added the new DIS processes that appear for
non-vanishing $Q^2$, when one photon is direct and the other
resolved, i.e. VMD or GVMD. For simplicity, first assume that one 
of the two photons is real, $Q_2^2=0$. For large $Q_1^2$, this 
DIS contribution can be given a parton-model interpretation,
\begin{equation}
\sigma_{\mr{DIS}\times\mr{res}}^{\gast\ga}(Q_1^2) \simeq
\frac{4\pi^2\aem}{Q_1^2} \; F_2^{\ga}(x,Q_1^2) \simeq
\frac{4\pi^2\aem Q^2}{(Q_1^2 + m_{\rho}^2)^2}\sum_{\q} e_{\q}^2 \, 
\left\{ x  q^{\ga}(x, Q_1^2) + 
x \overline{q}^{\ga}(x,Q_1^2) \right\} \;.
\label{eq:F2ga}
\end{equation} 
Note that this is only the resolved part of 
$\sigma_{\mr{tot}}^{\gast\ga}$. The direct contribution from
$\gast\ga \to \q\qbar$ comes in addition, but can be neglected
in the leading-order definition of $F_2^{\ga}$. We will therefore
use parton distribution parameterizations for the resolved photon, 
like SaS 1D \cite{SaSpdf}, to define the
$\sigma_{\mr{DIS}\times\mr{res}}^{\gast\ga}(Q_1^2)$.
Then eq.~(\ref{eq:siggastptotsimple}) generalizes to
\begin{eqnarray}
\sigma_{\mr{tot}}^{\gast\ga} (W^2, Q_1^2) & = &
\sigma_{\mr{DIS}\times\mr{res}}^{\gast\ga} \; 
\exp \left( - \frac{\sigma_{\mr{dir}\times\mr{res}}^{\gast\ga}}%
{\sigma_{\mr{DIS}\times\mr{res}}^{\gast\ga}} \right) +
\sigma_{\mr{dir}\times\mr{res}}^{\gast\ga} 
\nonumber \\
 & & + \sigma_{\mr{res}\times\mr{dir}}^{\gast\ga} +
\sigma_{\mr{dir}\times\mr{dir}}^{\gast\ga} +
\left( \frac{W^2}{Q_1^2 + W^2} \right)^n \;
\sigma_{\mr{res}\times\mr{res}}^{\gast\ga}  \;.
\label{eq:siggastgatot}
\end{eqnarray}
The large-$x$ behaviour of a resolved photon does not
agree with that of the proton, but for simplicity we will
stay with $n=3$.

The generalization to both photons virtual then gives
\begin{eqnarray}
\sigma_\mr{tot}^{\gast\gast} (W^2, Q_1^2, Q_2^2) & = &
\sigma_{\mr{DIS}\times\mr{res}}^{\gast\gast} \; 
\exp \left( - \frac{\sigma_{\mr{dir}\times\mr{res}}^{\gast\gast}}%
{\sigma_{\mr{DIS}\times\mr{res}}^{\gast\gast}} \right) +
\sigma_{\mr{dir}\times\mr{res}}^{\gast\gast} 
\nonumber \\
 & & + \sigma_{\mr{res}\times\mr{DIS}}^{\gast\gast} \; 
\exp \left( - \frac{\sigma_{\mr{res}\times\mr{dir}}^{\gast\gast}}%
{\sigma_{\mr{res}\times\mr{DIS}}^{\gast\gast}} \right) +
\sigma_{\mr{res}\times\mr{dir}}^{\gast\gast}  
\nonumber \\
 & & + \sigma_{\mr{dir}\times\mr{dir}}^{\gast\gast}
+ \left( \frac{W^2}{Q_1^2 + Q_2^2 + W^2} \right)^n \;
\sigma_{\mr{res}\times\mr{res}}^{\gast\gast}   \;,
\label{eq:siggastgasttot}
\end{eqnarray}
where the choice of damping factor for the last term again
is a simple guess for an extension.
When $Q_1^2 \gg Q_2^2$ the expression for 
$\sigma_\mr{tot}^{\gast\gast} (W^2, Q_1^2, Q_2^2)$
can be related to the structure function of a virtual photon,
$F_2^{\gast}(x,Q^2=Q_1^2,P^2=Q_2^2)$, where 
$x = Q_1^2/(Q_1^2 + Q_2^2 + W^2)$.

A comment about the exponential factors suppressing the 
DIS terms. Properly, each VMD/GVMD state should come with 
its ratio of direct to DIS cross sections. However, a number
of common factors divide out in this ratio: obviously the
probability to fluctuate to the state in question, but partly
also the form of the parton distributions in the state.
Therefore, it is a good approximation to define a common
exponential form for all VMD/GVMD state, based on the 
weighted average, which is equivalent to using the full
resolved term in both numerator and denominator. 

\subsection{Other Model Aspects}

In the preceding section we have mainly described the model as 
differential in $W$ (invariant mass of the $\gast\p$ or $\gast\gast$ 
system) and the hardest scales involved: $\pT$, $\kTi$ and $Q_i^2$. 
The discussion about the different processes has been limited 
to the parton-level only, but it is necessary to add parton 
showers, beam remnants, hadronization etc. in order to simulate 
complete events. The model is implemented in the \Py\ 6.151 event 
generator~\cite{pythia}, which includes the above and many other 
aspects, and it will be used in the following for the results 
presented. Clearly, here we can only give a few typical examples; 
with the help of the generator it is possible to study any specific 
experimental conditions and observables.

While based on the formalism outlined above, the complete
event simulation also involves some more model-specific
assumptions. We here present a few of those, to try to complement
the overall picture. Further refinements are possible in many 
places, and some of the known shortcomings are mentioned. 

The extension from $\p\p$ to $\gast\p$ and $\gast\gast$ collisions
requires two new aspects to be introduced in the event simulation
structure. One is the need to construct the $\e \to \e\gast$
kinematics, and to let the collision description depend not only on
the resulting $W$ but also on the selected $Q^2$ virtuality scale(s) 
of the photon(s). This part was described in~\cite{lutp9911}.

The other is the necessity to mix all the different reaction processes.
The main classification is here into 4 components for $\gast\p$ ---
VMD, GVMD, direct and DIS --- and 13 for $\gast\gast$ --- 4-squared
except that there is no DIS process on a direct photon or another 
DIS photon. Each of these components has a set of allowed subprocesses,
e.g. $\gast \q \to \q \g$ and $\gast \g \to \q \qbar$ for direct processes.
This set is the same for a VMD and a GVMD photon, but the two are
distinguished by the different parton distributions and total cross
section parameterizations used. In the Monte Carlo, a choice is first
made between the 4/13 components and then, for that component, among
the allowed subprocesses, according to maxima estimated at the 
initialization. Once the kinematics of the event has been fully selected,
the ratio of the actual cross section to the assumed one is used to
retain the event or to select a new component and subprocess.

The separation of VMD from GVMD requires access to parton distributions 
where those two components are explicitly made available separately.
Effectively this limits the choice to the SaS 
parameterizations~\cite{SaSpdf}, 
and with the assumed $k_0 \approx 0.5$ GeV to SaS 1D. 
An almost equivalent formalism could have been constructed in terms of 
a common resolved class, so that we could have gone from 4/13 to 3/6 
categories. While the calculation of partonic processes would have been 
simplified, aspects related to total cross section processes (elastic,
diffractive, low-$\pT$, etc.) would not, so we have not yet tried to
construct such a complete alternative machinery.

A VMD photon may be classified either as $\rho^0$, $\omega$, $\phi$
or $\Jpsi$. The parameterizations of total, elastic, single and double 
diffractive cross sections for $V\p$ and $V_1 V_2$ collisions of real 
photons are given in~\cite{SaSgaga}.
The respective inelastic non-diffractive cross section is obtained by 
subtraction, and sets the envelope within which the jet cross section 
is eikonalized. The events are ultimately classified, either by the 
hardest interaction that occurs, or as a low-$\pT$ event if there are 
no hard interactions. For virtual photons, both the total, elastic, 
single and double diffractive cross sections are dampened by the
same dipole factors, cf. eq.~(\ref{sigmatotVMDpvirt}). The jet cross section 
is obtained by the virtuality-dependent parton distributions, however,
so in the eikonalization procedure the mixture between jet and 
low-$\pT$ events is explicitly $Q^2$-dependent. For technical reasons, 
currently it is not possible to mix different dipole factors
for the different VMD states, so the $\rho/\omega$ mass is used 
throughout, thus giving a too fast dampening of the (rare) $\Jpsi$
events.

For the GVMD states, the mass selection is based on the identification
$m \simeq 2\kT$ and a $\kT$ spectrum e.g. as implied by 
eq.~(\ref{eq:sigmatotAnopTL}). Thus, neglecting the longitudinal-photon 
factor, one has a mass spectrum like $\d m^2/(m^2 + Q^2)^2$ in the range 
$2 k_0 < m < 2 \pTmin(W^2)$. This spectrum begins at 1~GeV, i.e.
a bit above $m_{\rho}$. Again there is not yet any provision for 
heavier quarks in the dipole dampening formulae, but the mass of an
$\s\sbar$ state is shifted by $m_{\phi} - m_{\rho}$ and that of
an $\c\cbar$ by $m_{\Jpsi} - m_{\rho}$, to ensure that the GVMD
spectrum starts above the lowest-lying state. The mass selected above 
then becomes the final-state mass of an elastically scattered GVMD 
particle. A diffractively scattered GVMD is assigned an excited mass 
according to a spectrum stretching from the selected mass and upwards
to the kinematical limit, essentially like $\d m^2/m^2$, as described 
in~\cite{SaSgaga}.

An elastic or diffractive GVMD state is in reality no different from 
a diffractive VMD state of the same mass. A low-mass system is allowed 
to decay to two hadrons, whereas more massive ones are considered
as strings stretched along the event axis, modulo limited 
transverse-momentum fluctuations. It is assumed that
a simple string is stretched between two endpoint quarks half
of the time, and a hairpin string arrangement with a gluon pulled
back the other half. Clearly this mixture is only a simple first
approximation to what is likely to be a more complex structure
\cite{IngSch}.

The mass of the GVMD state does not enter the description of 
inelastic non-diffractive events. However, here the $\kT \simeq m/2$
does provide a `primordial $\kT$' kick that can be transmitted
to the parton of a hard scattering, not only for GVMD$\times\p$ but
also in processes such as DIS$\times$GVMD. The corresponding number
ought to be $\kT \approx m_{\rho}/2$ for VMD and 
$\kT \approx m_{\p}/3 \approx m_{\rho}/2$ for a proton. However, in 
hadronic collisions much higher numbers than that are often required
to describe data, typically of the order of 1 GeV \cite{primkT} if
a Gaussian parameterization is used. Thus, an interpretation as a 
purely nonperturbative motion inside a hadron is difficult to maintain. 

Instead a likely culprit is the 
initial-state shower algorithm. This is set up to cover the region of
hard emissions, but may miss out on some of the softer activity,
which inherently borders on nonperturbative physics. By default, the
shower does not evolve down to scales below $Q_0 = 1$~GeV; if the
$\kT \simeq m/2$ scale of a GVMD photon is above this the evolution
is stopped already at the larger scale. (Whether $Q_0$ should best be
compared with $m$ or $m/2$ is an open issue.) Any shortfall in shower
activity around or below this cutoff then has to be compensated by the 
primordial $\kT$ source, which thereby largely loses its original meaning.

It could be argued that, while a VMD photon should have the same 
primordial $\kT$ spectrum as a hadron, a GVMD one should receive its
main contribution from the perturbative $\kT \simeq m/2$ contribution,
i.e. have a shape like $\d \kT^2/(\kT^2 + Q^2/4)^2$ in the range
$k_0 < \kT < \pTmin(W^2)$. However, this has the questionable consequence
that a low-mass GVMD state would have a smaller average `primordial $\kT$' 
than a VMD one. While we have retained such a possibility as an option,
our default has instead been to go to the other extreme, where the partons
in a GVMD state has a $\kT$ given by the vector sum of the perturbative
power spectrum and the same nonperturbative Gaussian smearing as for 
hadrons. Comparisons with data on photon remnant jets should eventually
shed more light on the appropriate procedure.

In most processes, initial- and final-state shower activity is routinely 
added to the lowest-order process, thereby providing an approximation to
higher-order QCD corrections. An exception is the DIS process 
$\gast\q \to \q$, where currently only final-state radiation has been
implemented. The technical reason why the initial-state radiation 
algorithm does not work here is that it is based on a definition of the $z$ 
of the splitting kernel as being the fraction by which the $\hat{s}$ of the
hard subprocess is reduced by an emission, and $\hat{s}=0$ in the DIS 
process above. The traditional $z$ interpretation is sensible if one
remembers that the complete DIS hard process is $\e\q \to \e\q$ rather
than only $\gast\q \to \q$. However, since we have split off the photon 
flux, the complete process is not trivially available currently. 
For small $Q^2$ values, this is no major problem, since events with 
activity above this scale belong to the direct event class. Therefore 
the shower is constrained only to populate the region below $Q^2$. The 
problem may become more severe in the high-$Q^2$ domain, where DIS 
processes provide the dominant event class and there is a large phase
space for shower activity. That is not the region of main interest to us 
here, however. 

The choice of a formalism, with the virtual photon flux separated from the 
hard processes, is there for two reasons. One is that it makes it possible 
to set up selection criteria for an event sample, e.g. on $W$ and $Q^2$
or in terms of scattered electron energies and angles, that are consistent 
across the different photon components. Another is that many nonperturbative
physics aspects, such as VMD total cross sections, best are formulated
in such terms.

Also a few other areas are not fully developed. One example is multiple
interactions, where only the simpler impact-parameter-independent option
can be used \cite{multint}, at least until one has constructed a model of
the spatial distribution of partons in a GVMD state. On the positive side,
options are available for several of the other aspects discussed in this
article, so that one may study the sensitivity to the assumptions made.

\section{Results}
\label{sec:Results}

In this section, some examples of event properties will be given,
with emphasis on the differences between the direct, VMD and 
anomalous/GVMD component of the photon. 
We will concentrate on a few distributions, starting in the 
photoproduction limit, and then explore how they gradually turn 
over from being dominated by the resolved processes to finally being
dominated by the bare direct and DIS ones as the photon virtuality 
increases. 

Owing to the resolved components of the photon, $\p\p$, $\ga\p$ and $\ga\ga$
events show similar behaviour for several event properties, such
as multiplicity and transverse energy flow. But clearly, 
differences are expected due to, for example, the additional 
direct event class in $\ga\ga$.

\subsection{Total Cross Sections}

\subsubsection{$\gast\p$ Total Cross Section}

\begin{figure} [t]
   \begin{center}
   \mbox{\psfig{figure=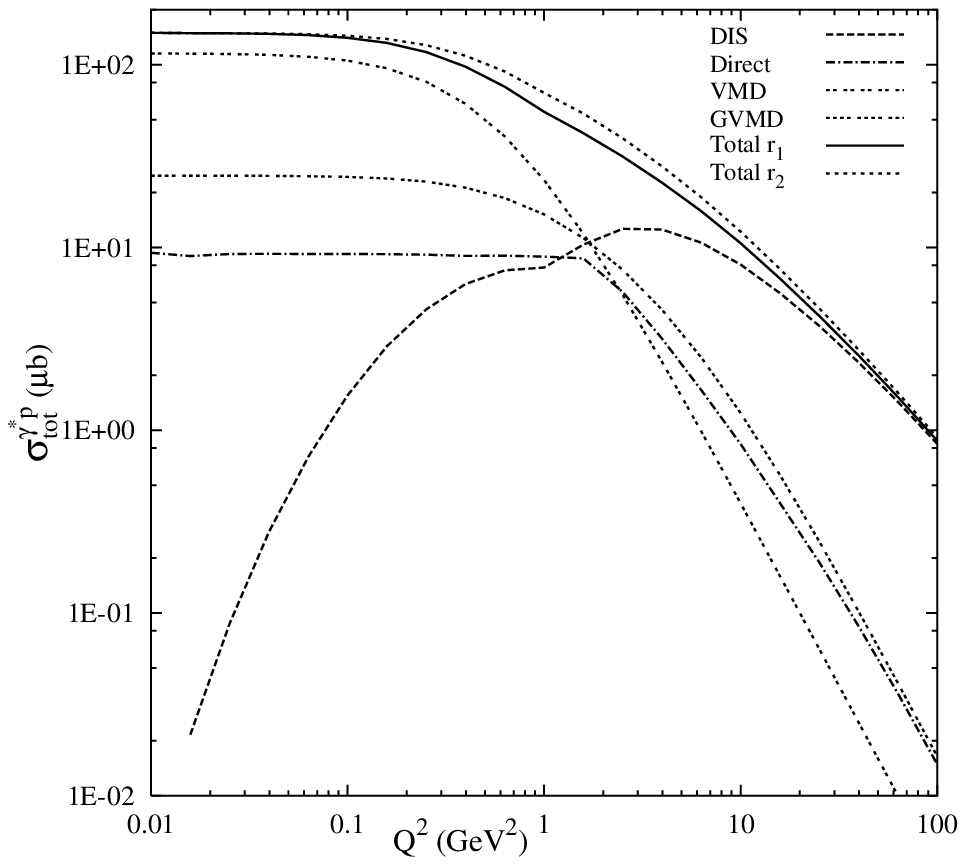,width=105mm}\hspace{-25mm}
	\psfig{figure=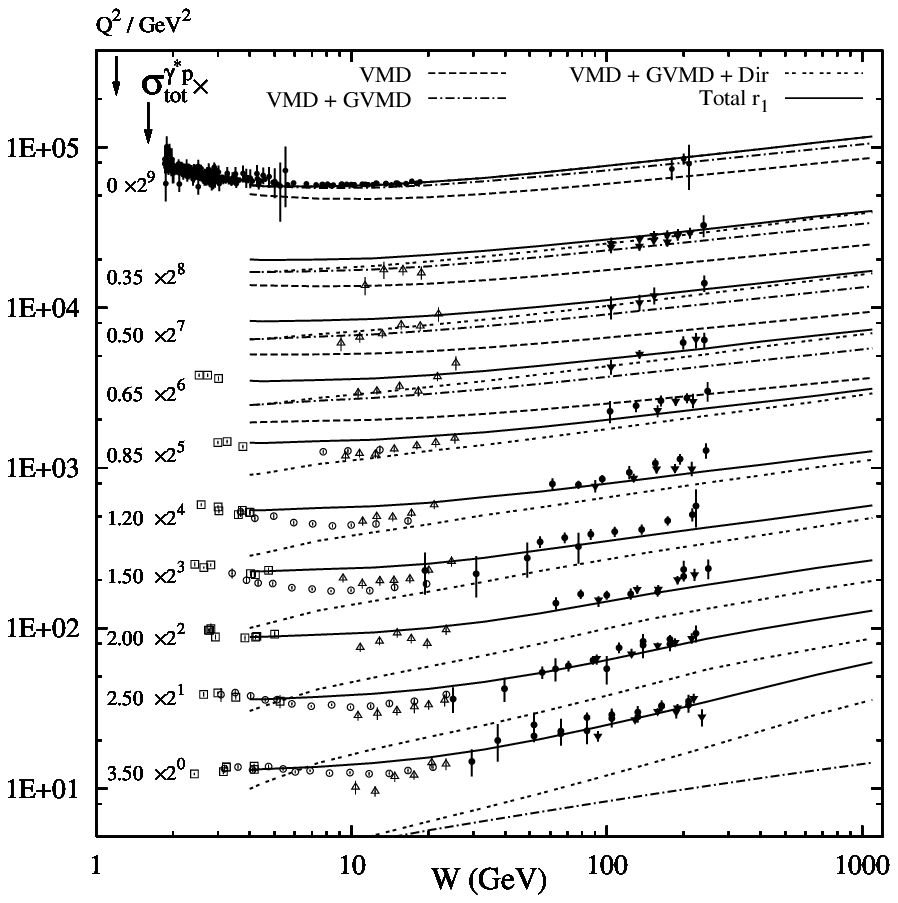,width=105mm}}
   \end{center}
\vspace{-5mm}\hspace{2cm}a)\hspace{8cm}b)\\
\captive{%
a) The total $\sigma^{\gast\p}_\mr{tot}$ cross section as function
of the photon virtuality at $W_{\gast\p}=100~\mr{GeV}$. The resolved
components shown are with alternative $r_1$ for the longitudinal 
contribution. 
b) Contributions to the total $\sigma^{\gast\p}_\mr{tot}$ cross 
section from the different event classes: direct, VMD, GVMD and DIS. 
For simplicity, the 
VMD and VMD+GVMD contributions are not shown for the lines 
corresponding to $Q^2$ values in the range $0.85-2.5~\mr{GeV}^2$.
See text for details. The legend for the data points can be found in 
Fig.~\ref{fig:ga*ptot}a.
\label{fig:ga*p}}
\end{figure}

The SaS~1D photon parton distribution and the CTEQ5L proton parton 
distribution will be used throughout in this section if not otherwise 
stated~\cite{SaSpdf,CTEQ5L}.
In Fig.~\ref{fig:ga*p}a, the $\gast\p$ cross section at 
$W_{\gast\p}=100~\mr{GeV}$ is shown as a function of the photon virtuality 
$Q^2$. The VMD and GVMD components contain contributions from 
resolved longitudinal photons with the $r_1$ alternative. An implicit
$y$ dependence enters the $r_i$ factors, 
eq.~(\ref{eq:rV1}) and~(\ref{eq:rV2}), to compensate
for the difference in photon flux between longitudinally and transversely 
polarized photons. 
To estimate the longitudinal contribution to the resolved component
in $\gast\p$ cross sections, a fixed $y$ is chosen in order to set 
the relative amount of transverse and longitudinal photons. The
value of $y$ consequently depends on the beam energy, and at HERA 
$W_{\gast\p}=100~\mr{GeV}$ would correspond to $y \simeq 0.1$, a value which 
will be used in the following. The parameter $a$, appearing
in $r_i$, is unknown and need to be determined by data. As a starting point 
for the discussion, $a$ is chosen to 0.5.

In the photoproduction limit, the VMD component dominates the
total cross section whereas the DIS one has a vanishing cross section
by construction, see Fig.~\ref{fig:ga*p}a. The GVMD and direct processes 
share the remaining part, about 20\%, of the total cross section.
It follows from the introduction of the dipole factors in 
eq.~(\ref{sigmatotVMDpvirt}) that, when considering total $\gast\p$ cross 
sections, the VMD class is the dominating component up to photon 
virtualities of the order of a vector meson mass. Similarly, the 
characteristic virtualities for the GVMD class to still be important are 
of the order of the $k_1$ parameter discussed in section~\ref{photop}.

The direct processes are simulated in the region where the hard scale $\kT$
of the photon--parton scattering fulfill $\kT > \max(k_1, Q)$, 
cf. Fig.~\ref{fig:classDIS}. At low $Q^2$ values, the 
\mbox{$k_1=\pTmin(W^2) \simeq 1.3~\mr{GeV}$} provides a fix limit on the 
available phase space, but at higher  $Q^2$ the phase space shrinks,
explaining the kink in Fig.~\ref{fig:ga*p}a for the direct 
cross section at $Q^2=k_1^2$. So it is rather this condition that drives the
dampening at large $Q^2$ than the explicit $Q^2$ dependence appearing in the
matrix elements. 
A kink can also be seen in the DIS cross section; it is an artifact from 
the freezing of the parton distributions below their range of applicability
(as introduced in section~\ref{sec:DIS}). Hence, the CTEQ5L proton parton 
distribution has $Q^2_\mr{min}=1~\mr{GeV}$. The factors introduced for the
DIS component, eq.~(\ref{eq:F2mod}), together with the exponential dampening, 
eq.~(\ref{eq:LODISmod}), make the unusual form of the DIS cross section ---
to be discussed in more detail below.
  
\begin{figure} [t]
   \begin{center}
   \mbox{\psfig{figure=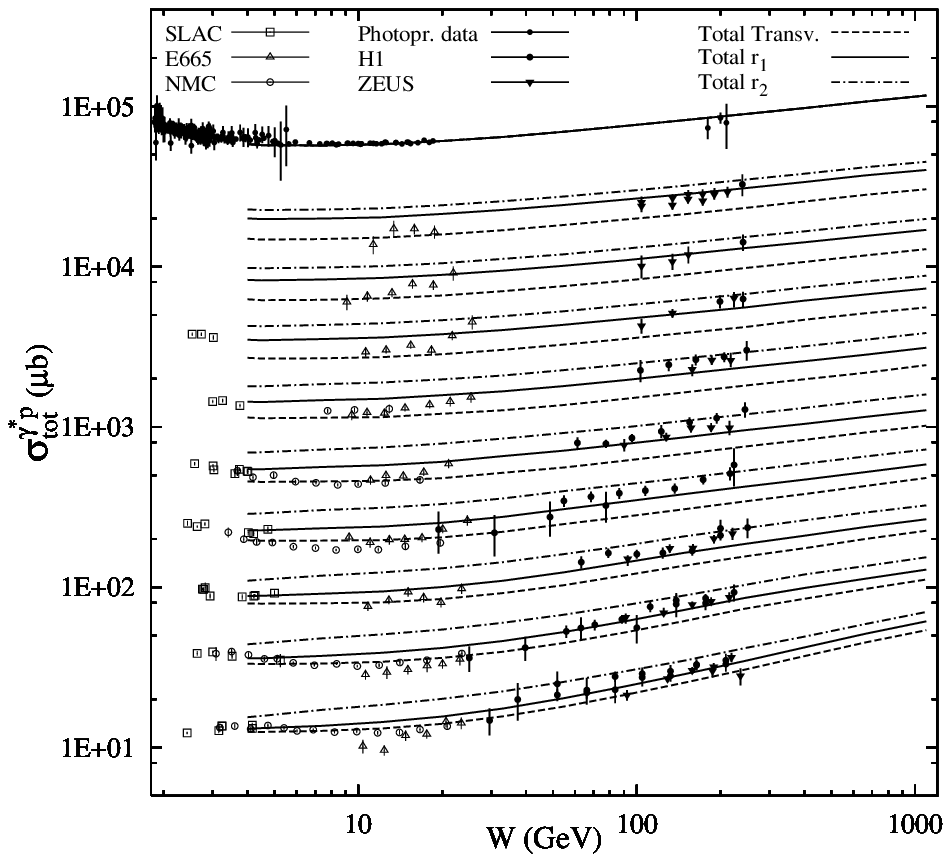,width=105mm}\hspace{-25mm}
	\psfig{figure=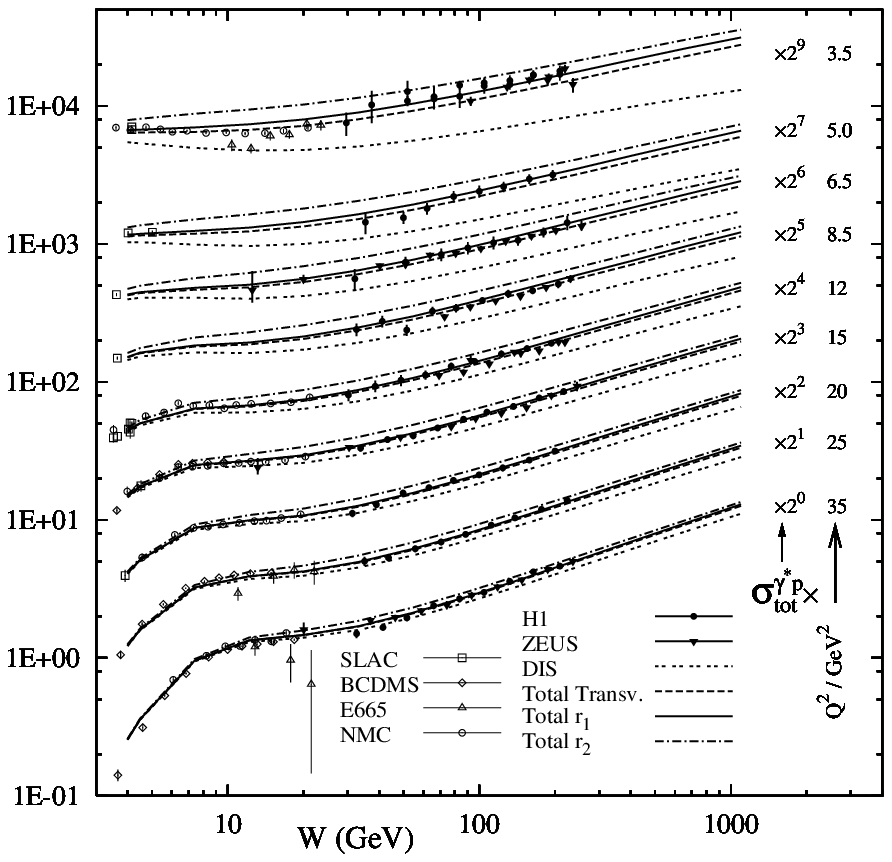,width=105mm}}
   \end{center}
\vspace{-5mm}\hspace{2cm}a)\hspace{8cm}b)\\
\captive{
a) The total contribution to $\sigma^{\gast\p}_\mr{tot}$ 
from the different event classes. The two different longitudinal 
alternatives are compared to the case with only transverse photons 
considered. The data and `Total $r_1$' is the same as in
 Fig.~\ref{fig:ga*p}b.
b) Same as in a but for a different
$Q^2$ range and also the DIS contribution is shown. 
For reference, the $Q^2=3.5~\mr{GeV}^2$ lines are the same, however.
\label{fig:ga*ptot}}
\end{figure}

The proton structure function $F_2$ has been measured at various
$x$ and $Q^2$ values \cite{F2data} and is related to the total
virtual photon--proton cross section $\sigma^{\gast\p}_\mr{tot}$
through eq.~(\ref{eq:F2}). In Fig.~\ref{fig:ga*p}b and~\ref{fig:ga*ptot}a, 
$\sigma^{\gast\p}_\mr{tot}$ is shown from zero $Q^2$,
the total photoproduction cross section~\cite{photopdata}, to medium $Q^2$ 
values as a function of the invariant mass $W$ of the 
$\gast\p$-system. 
Results from the model are compared with data from different 
fixed-target and HERA experiments. The contribution from the different 
event classes: direct, VMD, GVMD and DIS are shown in
Fig.~\ref{fig:ga*p}b, and sum up to the total contribution as given in 
eq.~(\ref{eq:siggastptotsimple}). 
The DIS contribution can be 
obtained by subtracting the `VMD+GVMD+Dir' contribution from the 
`Total r$_1$' one. 
With increasing photon virtuality, the energy dependence of $k_1$ makes 
the GVMD states to be less dampened at large energies as compared to low 
ones. This is also indicated in Fig.~\ref{fig:ga*p}b by comparing the 
`VMD' and `VMD+GVMD' lines at different $Q^2$ and $W$. 

In Fig.~\ref{fig:ga*ptot}a, the total contribution obtained with different
assumptions of the longitudinal contribution for resolved photons are 
compared to the case with transverse photons only. With the other model 
assumptions made, a non-vanishing longitudinal contribution is 
indicated. The alternative with $r_1$ gives a good fit to data, 
but overshoots at low $W$ values for some $Q^2$ values.
In the high $Q^2$ region, showed in Fig.~\ref{fig:ga*p}a and at the bottom 
in Fig.~\ref{fig:ga*ptot}b, the choice of longitudinal alternative for 
the resolved component is irrelevant for the total cross section since 
the resolved event classes give a 
negligible contribution. 
Hence, the parameter $a$ should be tuned at intermediate $Q^2$ values. 
At lower $Q^2$ values, however,
the $r_2$ alternative is overshooting data significantly. 
Taking $a=0.5$ as a satisfactory value for the $r_1$ alternative, 
the $r_2$ one require $a=0.2$ to give an equally good description (not shown). 

The direct processes increase with $W_{\gast\p}$ faster than the DIS one, 
and consequently also the exponential suppression of the DIS term. 
Therefore they contribute substantially to the total cross section in 
the large $W_{\gast\p}$ (small $x$) region, even at photon virtualities 
up to tens of GeV$^2$, shown for $W_{\gast\p}=100~\mr{GeV}$ in 
Fig.~\ref{fig:ga*p}a.
The DIS and direct classes are often combined and clearly dominate the 
total cross section at $Q^2=3.5~\mr{GeV}^2$ which is above the 
$m_V^2$ and $k_1^2$ scales of the VMD and GVMD states discussed above.
Due to the negligible contribution from the direct processes
at $W_{\gast\p}=10~\mr{GeV}$, the exponential suppression factor 
for the DIS term is close to unity --- independently of $Q^2$.
Hence, the sum of the direct and DIS contributions reproduce the 
$\sigma_{\mr{DIS}}^{\gast\p}$ cross section, eq.~(\ref{eq:DIScomp}).
For $W_{\gast\p}=100$ and $1000~\mr{GeV}$ this is the case for 
$Q^2 > 5$ and $8~\mr{GeV}^2$, respectively. 

\subsubsection{$\gast\ga$ Total Cross Section}

\begin{figure} [t]
   \begin{center}
   \mbox{\psfig{figure=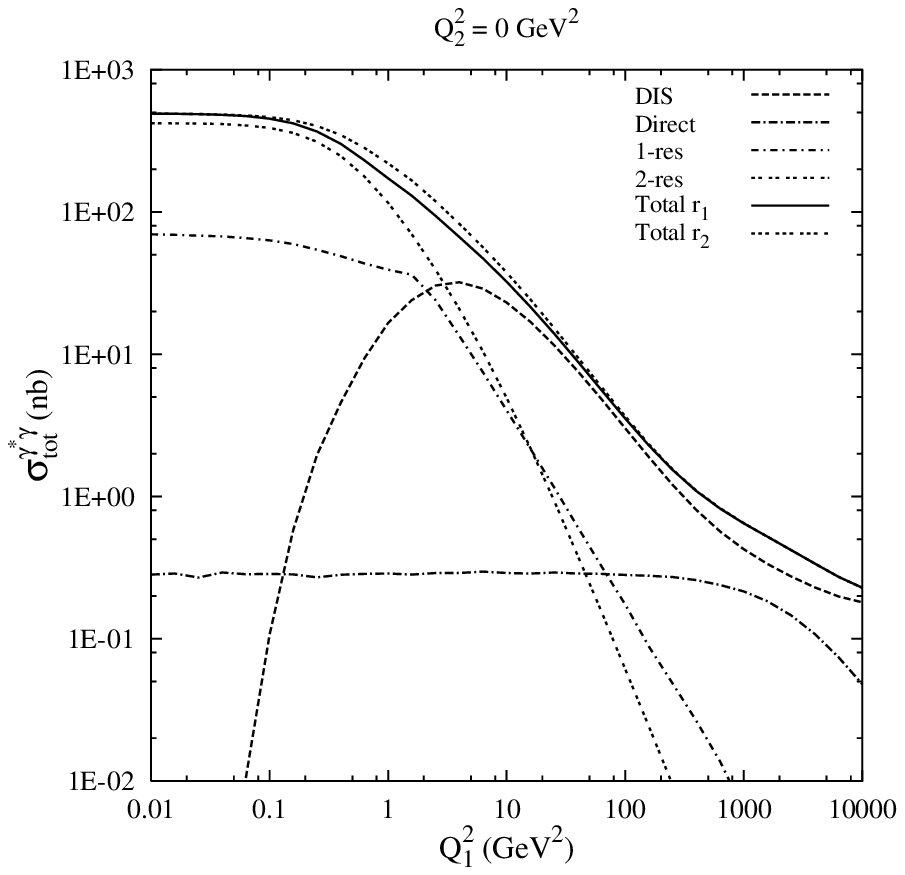,width=105mm}\hspace{-25mm}
	\psfig{figure=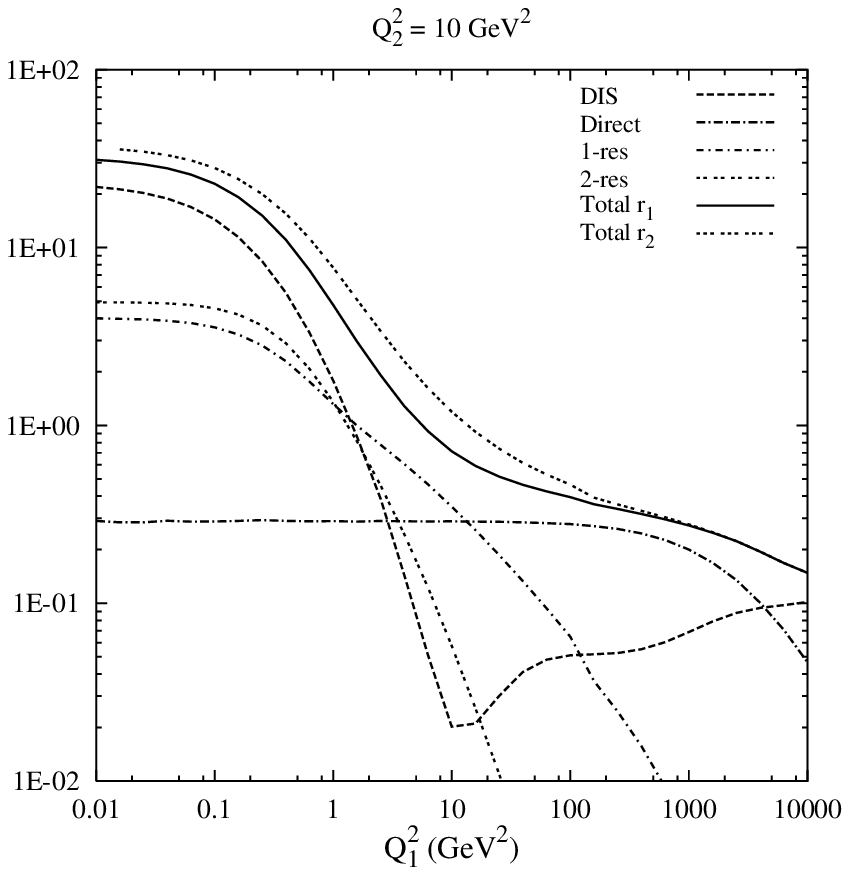,width=105mm}}
   \end{center}
\vspace{-5mm}\hspace{2cm}a)\hspace{8cm}b)\\
\captive{%
a) The $\sigma^{\gast\ga}_\mr{tot}$ at $W_{\gast\ga}=100~\mr{GeV}$ 
as a function of the photon virtuality $Q_1^2$ of one of the photons 
is shown for the DIS, direct, single-resolved (1-res), double-resolved 
(2-res) and total contribution. The other photon is real, 
$Q_2^2=0~\mr{GeV}^2$.
b) Same as in a) but for $\gast\gast$ with a target photon virtuality 
$Q_2^2=10~\mr{GeV}^2$.
\label{fig:ga*gatotQ2}}
\end{figure}

For simplicity, a fixed $y_i=0.1$ is also used in photon--photon 
collisions for the calculation of the longitudinal resolved photon 
contributions. In Fig.~\ref{fig:ga*gatotQ2}a, the 
$\sigma^{\gast\ga}_\mr{tot}$ at $W_{\gast\ga}=100~\mr{GeV}$ 
is shown as a function of one of the photon virtualities; the other 
photon is real. The different 
components are shown separately and sum up to the total contribution with 
the $r_1$ alternative. Additionally, relative to the corresponding 
$\gast\p$ case, the direct cross section enters and starts to be 
significantly dampened only when $Q_1^2 \simeq W^2$. 
The kink seen for the single-resolved case can again
be explained by the constraint of the hard scale in the scattering 
process to be larger than the virtuality of the direct photon. 

The same distribution is shown for the case of two virtual photons in 
Fig.~\ref{fig:ga*gatotQ2}b. The low virtuality end of $Q_1^2$ of course
corresponds to the cross sections at $Q_1^2=10~\mr{GeV}^2$ in 
Fig.~\ref{fig:ga*gatotQ2}a. The DIS cross section dominates in this 
region, with the first photon as the target probed by the second one.  
This is interchanged for $Q_1^2 \geq Q_2^2$. It is noticeable that all
the different processes shown are of about the same importance when 
$Q_1^2$ is in the neighbourhood of $Q_2^2$.

\begin{figure} [t]
   \begin{center}
   \mbox{\psfig{figure=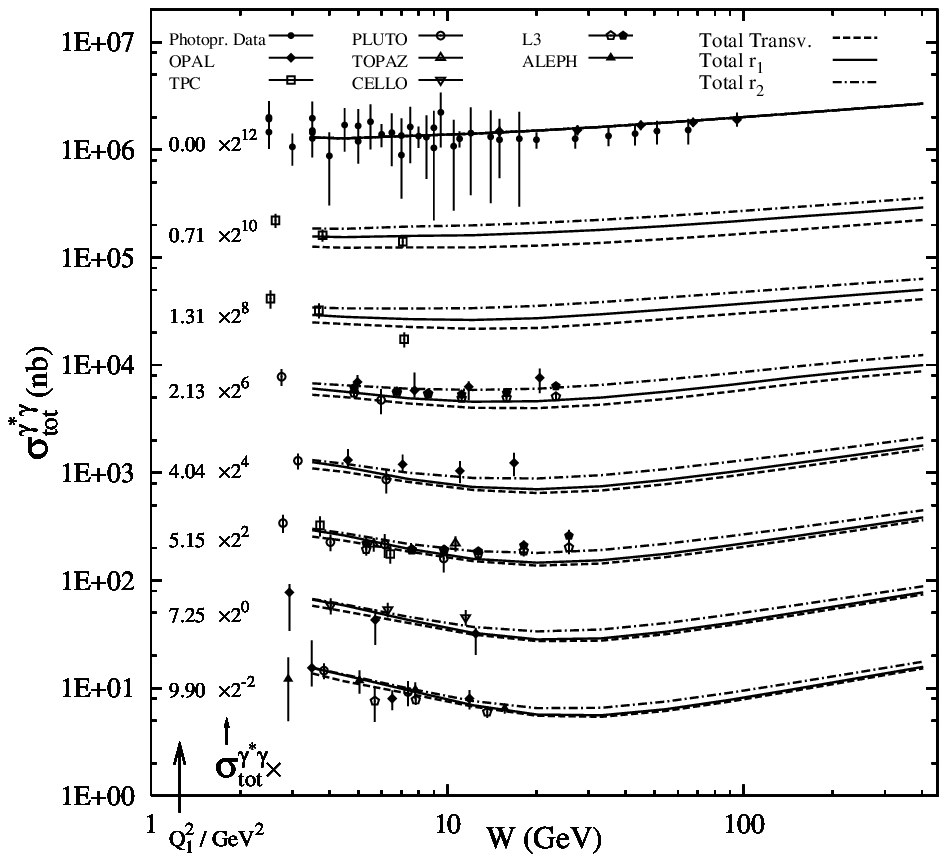,width=105mm}\hspace{-25mm}
	\psfig{figure=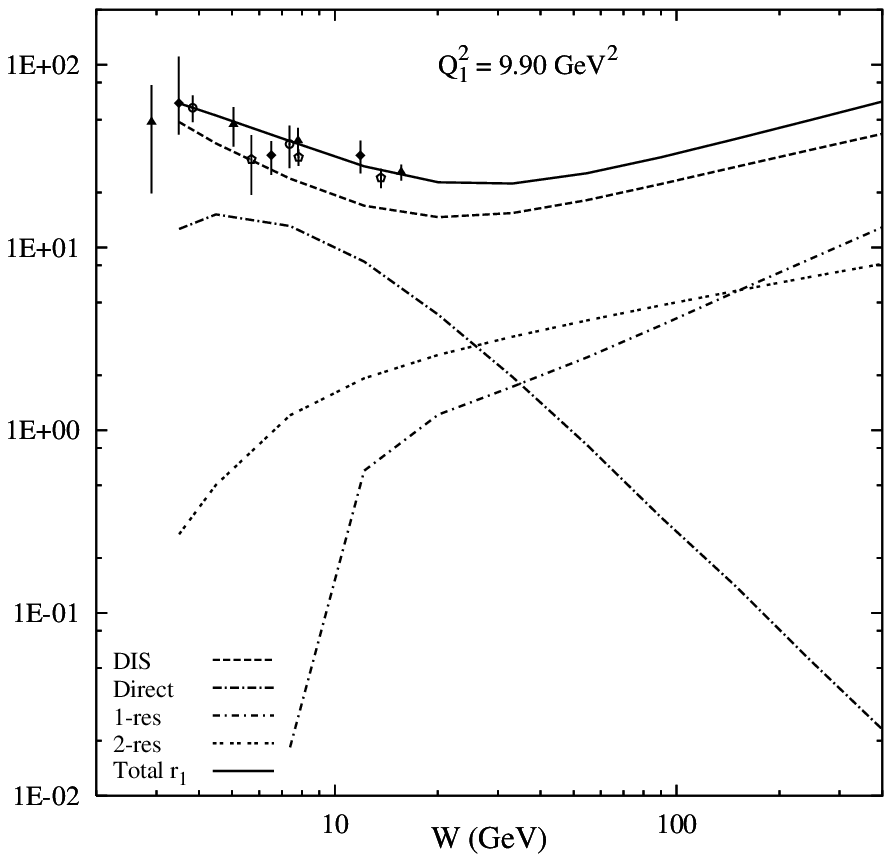,width=105mm}}
   \end{center}
\vspace{-5mm}\hspace{2cm}a)\hspace{8cm}b)\\
\captive{%
a) Same as in Fig.~\ref{fig:ga*ptot}b but for $\gast\ga$ instead of 
$\gast\p$.
b) The $\sigma^{\gast\ga}_\mr{tot}$ as a function of 
the invariant mass of the $\gast\ga$ system is shown for the DIS, direct,
single-resolved (1-res), double-resolved (2-res) and total contribution.
\label{fig:ga*gatot}}
\end{figure}


For various reasons, the photon $F_2^\ga$ measurements~\cite{F2gdata} 
are both less precise and available in a smaller kinematical 
range than in the proton case. 
In Fig.~\ref{fig:ga*gatot}a, the $\sigma^{\gast\ga}_\mr{tot}$ is shown
as a function of $W$ for different $Q^2$ values, similarly to the study 
in the previous section. With $a=0.5$ for both alternatives of the
longitudinally resolved photons, data does not discriminate between them.
The $r_1$ alternative is below the high-energy end data points
for the $Q^2=4.04$ and $5.15~\mr{GeV}^2$ lines   
but the errors and spread in data do not give an unambiguous conclusion.

Fig.~\ref{fig:ga*gatot}b shows the contribution to 
$\sigma^{\gast\ga}_\mr{tot}$ for some different event classes at 
$Q^2=9.90~\mr{GeV}^2$. The direct and DIS ones are dominating at low 
$\gast\ga$ invariant masses whereas the single-resolved ones increase in
importance at high $W_{\gast\ga}$. 
The direct processes show the characteristic $1/W^2$ fall 
off~\cite{lutp9911} and thereby give its major contribution to the 
region where the valence quarks dominate the contribution to $F_2^{\ga}$.
Again, the single-resolved processes (1-res; cf. direct ones in $\gast\p$) 
increase with energy, which therefore dampens the DIS term slightly at 
large $W_{\gast\ga}$, a region which is dominated by the gluon content of
the photon.
As resolved processes in $\gast\p$, double-resolved ones (2-res) in 
$\gast\ga$ increase with $W_{\gast\ga}$. 

\subsubsection{$\gast\gast$ Total Cross Section}

\begin{figure} [t]
   \begin{center}
   \mbox{\psfig{figure=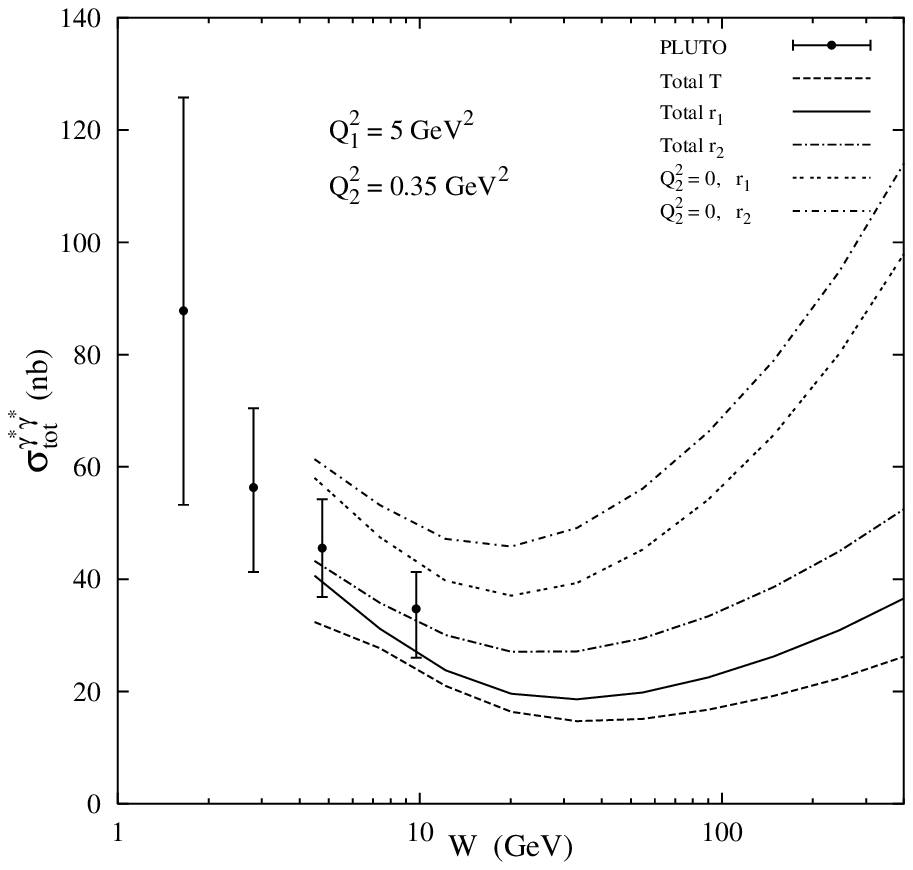,width=105mm}\hspace{-25mm}
	\psfig{figure=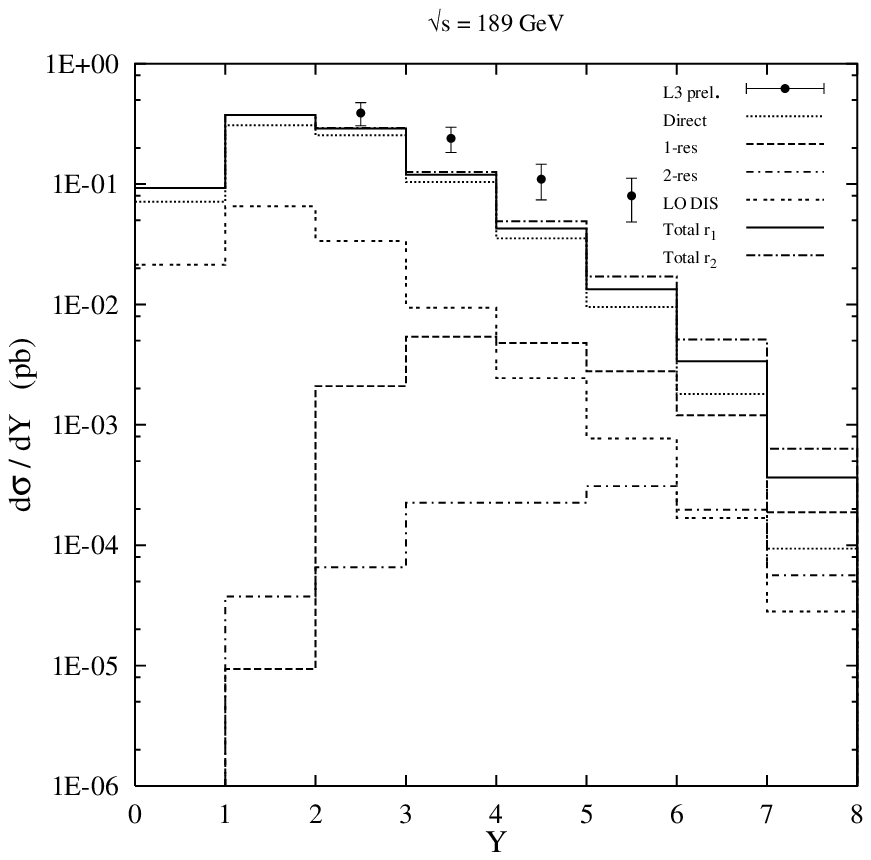,width=105mm}}
   \end{center}
\vspace{-5mm}\hspace{2cm}a)\hspace{8cm}b)\\
\captive{
a) The $\sigma^{\gast\gast}_\mr{tot}$ cross section as a function of 
$W_{\gast\gast}$ for a probing and target photon with virtualities 
5 and $0.35~\mr{GeV}^2$ respectively. The data points from PLUTO 
are obtained from the measured $F_\mr{eff}$ through eq.~(\ref{eq:F2}).
For reference, the results with a real photon target is shown, $Q_2^2=0$
b) The differential $\e^+\e^- \ra \e^+\e^- + \mr{hadrons}$ 
cross section as a function of $Y=\ln (y_1y_2s/\sqrt{Q_1^2Q_2^2})$. 
The different components shown add up to the `Total $r_1$'. 
\label{fig:L3}}
\end{figure}

There are hardly any systematic data on $\gast\gast$ cross sections.
Prior to the recent and ongoing LEP studies, essentially the only 
publication is the $F_\mr{eff}$ measurement by PLUTO~\cite{PLUTOg*g*}.
A comparison with these data points is found in Fig.~\ref{fig:L3}a.
The low $W$ values --- the first point is in the resonance region --- 
makes the comparison especially precarious, and the second photon
is also not all that virtual with its 
$\langle Q_2^2 \rangle = 0.35$~GeV$^2$. The main
point of the plot is thus not the acceptable agreement with data,
but to illustrate the amount of reduction of the cross section relative 
to the real-photon alternative $Q_2^2 = 0$ and the dependence on assumed 
longitudinal contributions.
 
Double-tag two-photon events, $\e^+\e^- \ra \e^+\e^- + \mr{hadrons}$, 
have been measured by the L3 collaboration~\cite{L3g*g*}.
The differential $\e^+\e^-$ cross section w.r.t. the variable
\begin{equation}
Y=\ln \left( \frac{y_1y_2s}{\sqrt{Q_1^2Q_2^2}} \right) 
\label{eq:Y}
\end{equation}
is shown in Fig.~\ref{fig:L3}b. The two scattered leptons are required to
be within the polar angle of $30<\theta_i<60~\mr{mrad}$ (with respect to 
the incoming beams) and to have an energy larger than 30~GeV. This plot 
is expected to provide a clean test of the BFKL behaviour, with a 
cross section increasing at large $Y$ \cite{BFKLtest}, but the data
does not support such a behaviour, or at least less of it than expected.
Our model agrees with the data at small $Y$, but tends to fall below at
larger values. This is almost entirely dictated by the drop in the
dominant direct contribution. The resolved contributions --- which are
the ones that could be used to represent a BFKL behaviour in our framework 
--- do come up at large $Y$, but nowhere near enough to make a significant
contribution. We remind that the resolved contributions are suppressed
in $Q^2$ by simple dipole factors; a less steep drop and thereby a larger
resolved contribution could be motivated e.g. if the eikonalization of the 
jet cross section is less significant for a virtual photon. Instead of the
conventional increase of the VMD cross section like $s^{\epsilon}$ with
$\epsilon \approx 0.08$, eq.~(\ref{sigmatotAB}), a larger effective value 
like $\epsilon \approx 0.2 - 0.3$ could thus easily be accommodated for
virtual photons \cite{Qdepeps}. For the moment,
however, we prefer to await further experimental data. In particular,
more detailed analyses of event properties could provide some insight on
the relative mixture of event classes at large $Y$.
 
\subsection{Event Properties}

\begin{figure} [tp]
   \begin{center}
   \mbox{\psfig{figure=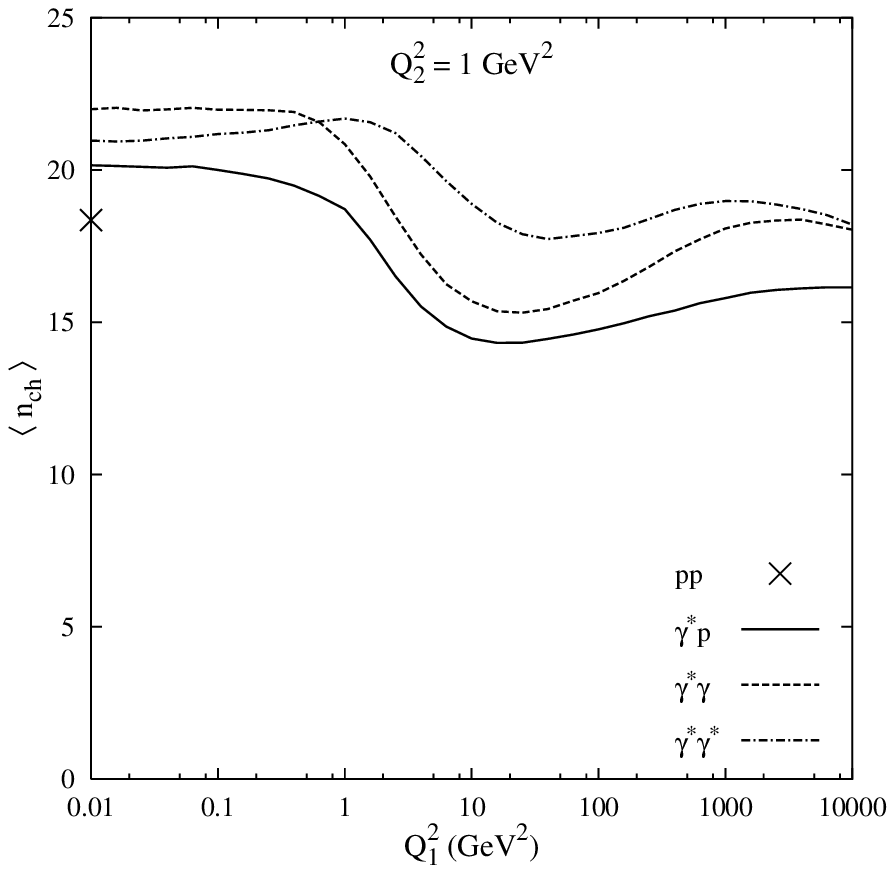,width=105mm}\hspace{-25mm}
	\psfig{figure=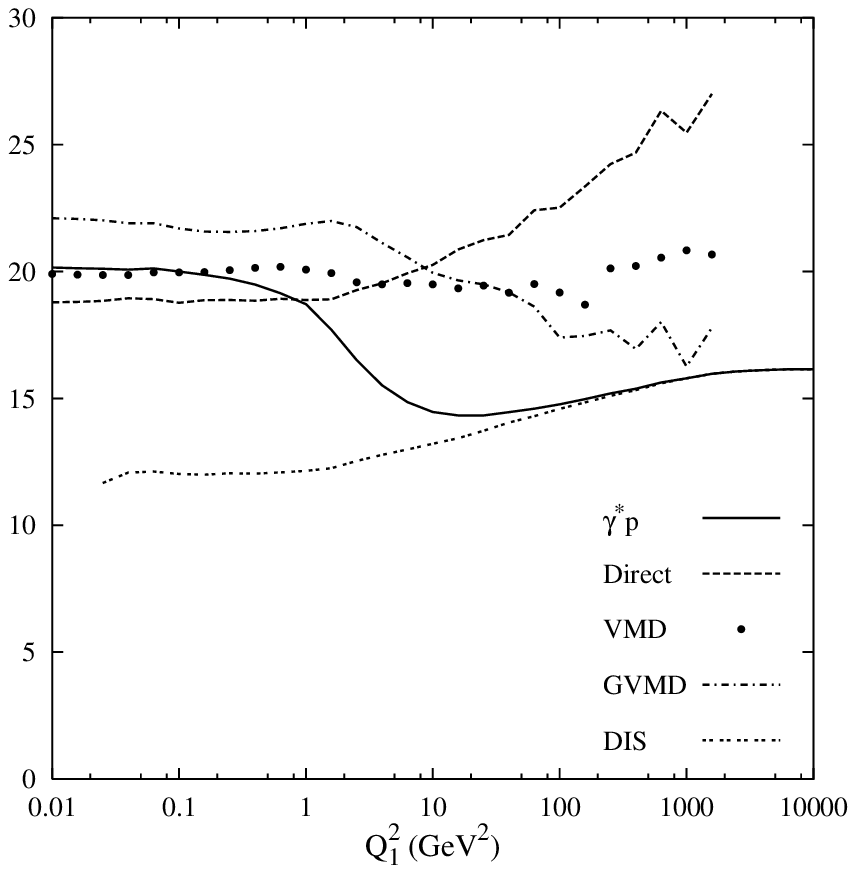,width=105mm}}
   \end{center}
\vspace{-5mm}\hspace{2cm}a)\hspace{8cm}b)
   \begin{center}
   \mbox{\psfig{figure=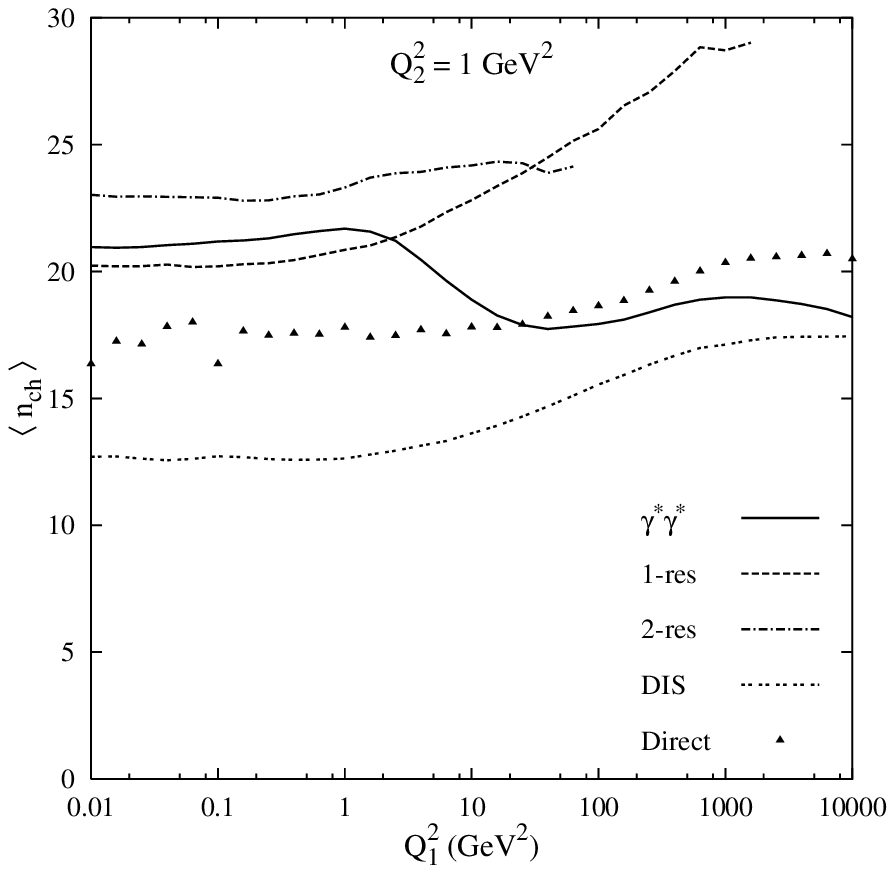,width=105mm}\hspace{-25mm}
	\psfig{figure=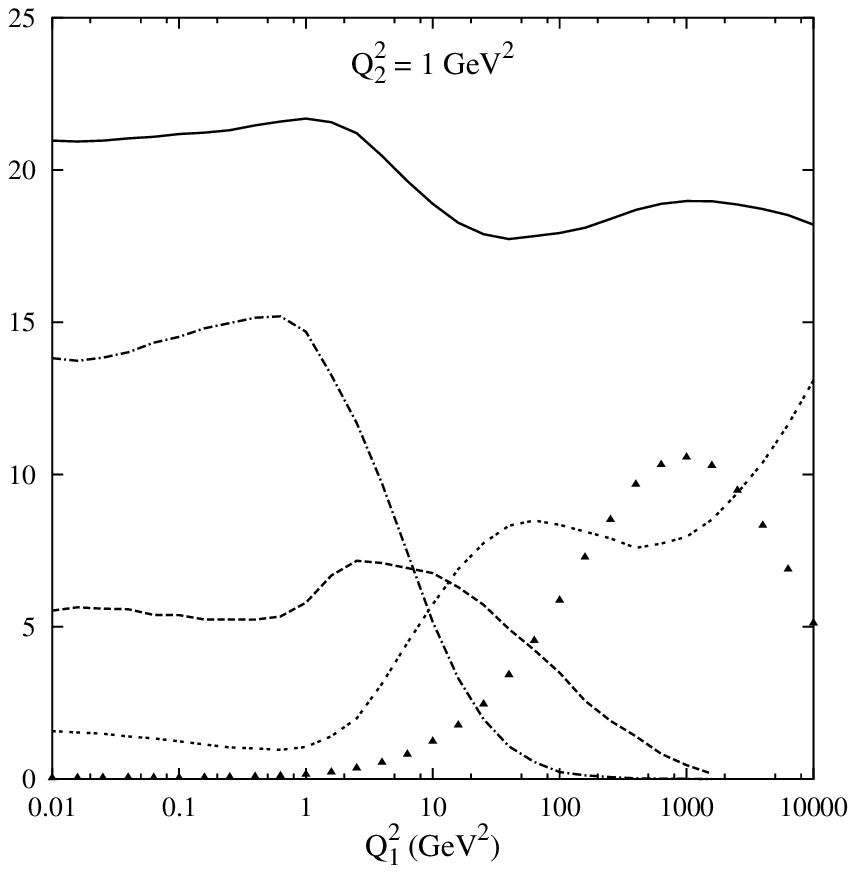,width=105mm}}
   \end{center}
\vspace{-5mm}\hspace{2cm}c)\hspace{8cm}d)\\
\captive{%
a) 
The average charged multiplicity $\langle n_\mr{ch} \rangle$ as a function 
of the photon virtuality for $\gast\p$, $\gast\ga$ and $\gast\gast$ events 
at a center of mass energy of 100~GeV. The result from $\p\p$ events is 
indicated with a cross on the $y$-axis. For the  $\gast\gast$ events, the 
other photon virtuality is kept fixed at $Q_2^2=1~\mr{GeV}^2$. 
(In $\gast\ga$ $Q_2^2=0~\mr{GeV}^2$.) No diffractive or elastic events are
considered.
b) The results from the different components in $\gast\p$, averaged over
the number of events of the respective kind. 
c) The result from different components in $\gast\gast$, averaged over
the number of events of the respective kind. 
d) As in c) but averaged over the total number of $\gast\gast$ events.
\label{fig:nch}}
\end{figure}

\begin{figure} [tp]
   \begin{center}
   \mbox{\psfig{figure=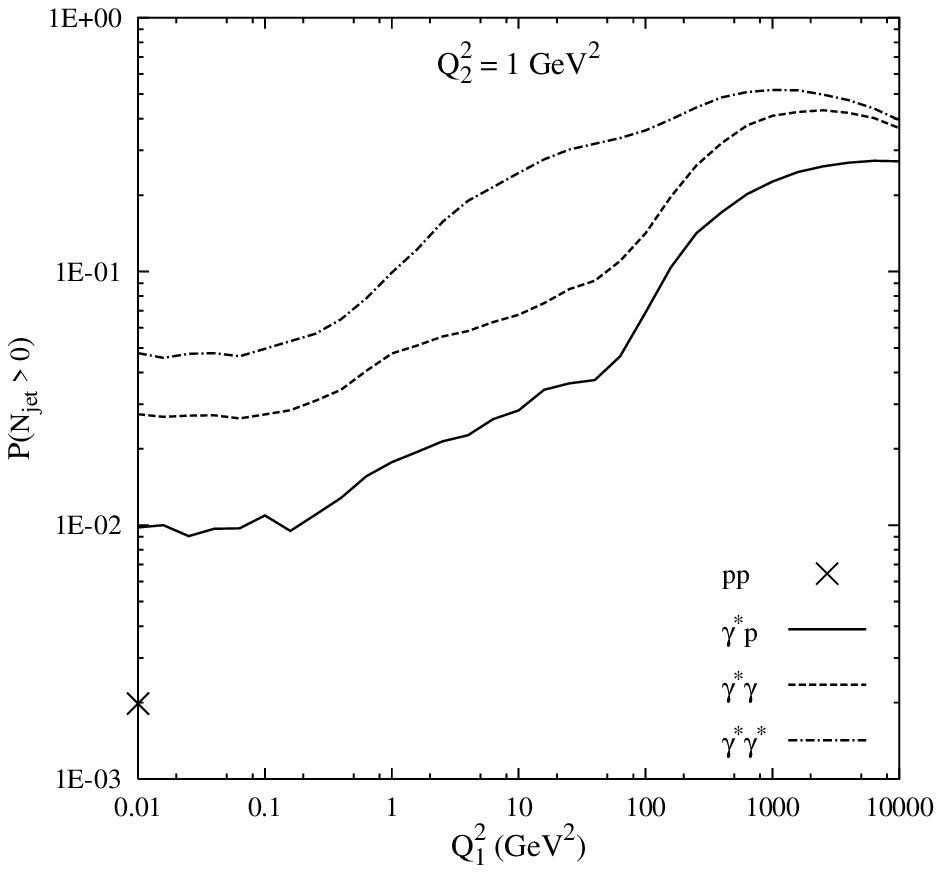,width=105mm}\hspace{-25mm}
	\psfig{figure=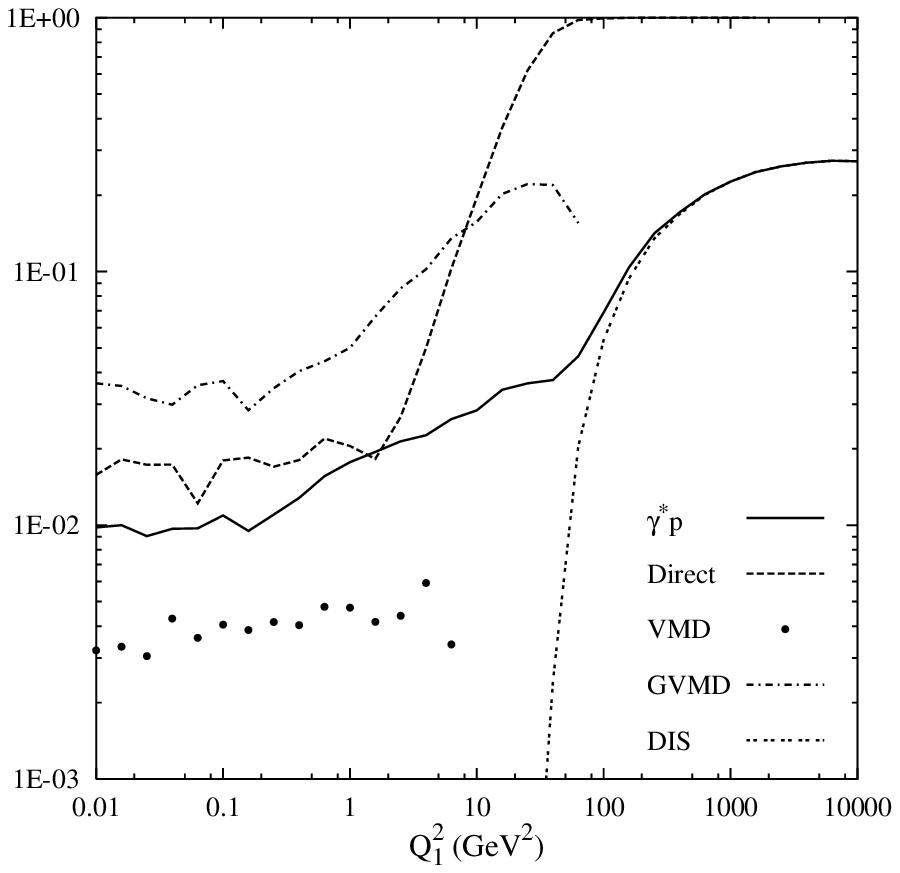,width=105mm}}
   \end{center}
\vspace{-5mm}\hspace{2cm}a)\hspace{8cm}b)
   \begin{center}
   \mbox{\psfig{figure=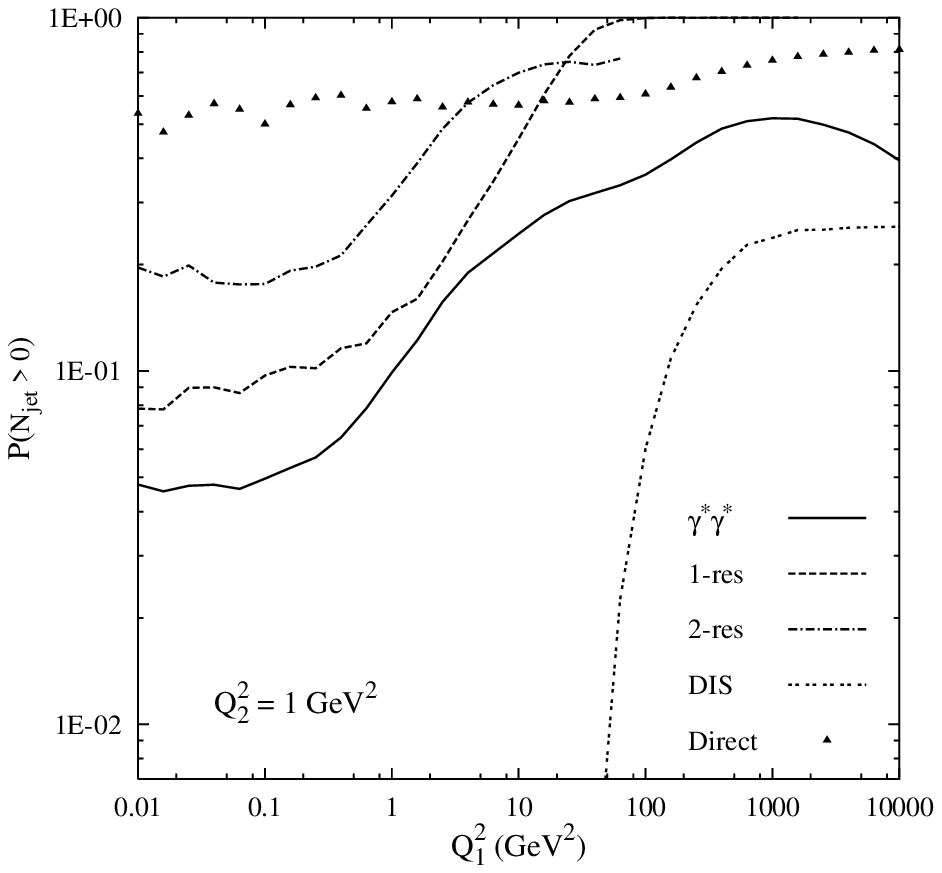,width=105mm}\hspace{-25mm}
	\psfig{figure=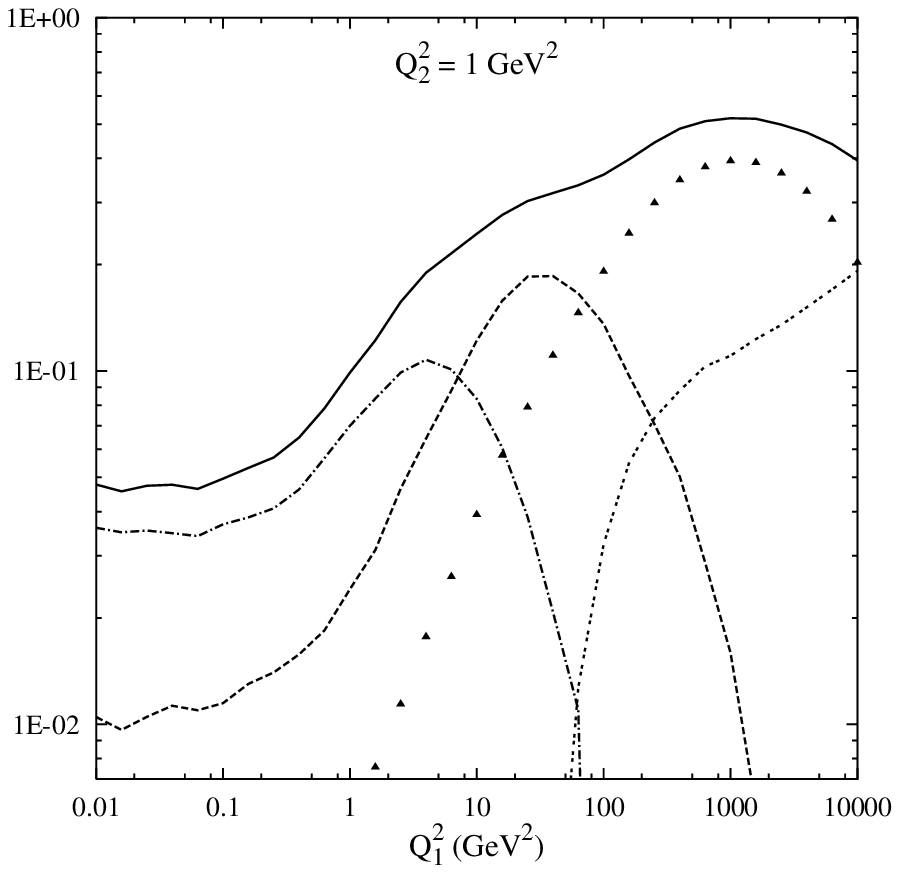,width=105mm}}
   \end{center}
\vspace{-5mm}\hspace{2cm}c)\hspace{8cm}d)
\captive{%
a) The fraction of events that contain at least one jet  with an 
$E_{\perp} > 5~\mr{GeV}$ inside a cone of radius 
$R = \sqrt{(\Delta y)^2+(\Delta\phi)^2} < 1$. $W=100~\mr{GeV}$ 
in all cases. No diffractive or elastic events are considered.
b) The results from the different components in $\gast\p$, averaged over
the number of events of the respective kind.
c) The result from different components in $\gast\gast$, averaged over
the number of events of the respective kind. 
d) As in c) but averaged over the total number of $\gast\gast$ events.
\label{fig:jetrate}}
\end{figure}

Some event properties are fairly similar between $\p\p$, $\gast\p$ and
$\gast\gast$, while others differ markedly. The charged 
multiplicity, Fig.~\ref{fig:nch}, is an example of the former, while
the jet rate, Fig.~\ref{fig:jetrate}, is of the latter.

Over a wide range of $Q^2$ values, the $\gast\p$ and $\gast\gast$ events
have average charged multiplicities within $\pm 20$\% of the $\p\p$ value,
Fig.~\ref{fig:nch}a. (Elastic and diffractive events have not been considered.
Their inclusion would reduce average multiplicities and change some details,
but leave a similar overall picture.) For $\gast\p$ at small $Q^2$, this is
a consequence of the dominance of VMD events, with the characteristic
features of hadronic physics, including such aspects as multiple
parton--parton interactions. The VMD multiplicity is fairly $Q^2$-independent,
Fig.~\ref{fig:nch}b. The drop to a somewhat lower $\gast\p$ multiplicity 
around $Q^2 \approx 1 - 10$ GeV$^2$ instead comes from the transition
from VMD dominance to DIS ditto, where the DIS events are cleaner by
consisting of only one string piece stretched directly between the kicked-out 
quark and the beam remnant. Remember that the DIS events here by definition 
are low-$\pT$ ones, with the high-$\pT$ part in the direct class, with a higher
multiplicity. As the borderline between the two shifts with $Q^2$, both
individually show an increasing multiplicity: the direct by corresponding  
to a smaller fraction of higher-$\pT$ events and the DIS by allowing an
increasing admixture of jet events. Figures \ref{fig:nch}c and
\ref{fig:nch}d show the corresponding $Q_1^2$ dependence of the 
$\gast\gast$ multiplicity, for a fixed $Q_2^2 = 1$~GeV$^2$, normalized to
the number of event of each kind and to the total number, respectively.
Again the disappearance of the doubly-resolved event class is responsible
for a drop in the multiplicity, with the DIS and direct processes taking 
over. We remind that in $\gast\gast$ the high-$\pT$ part of the DIS process is
the single-resolved one, corresponding to the direct one of $\gast\p$,
while the $\gast\gast$ direct class has no correspondence in $\gast\p$.

When it comes to the fraction of events that contain at least one jet,
the differences are orders of magnitude, both with a hierarchy of 
increasing jet fraction from $\p\p$ to $\gast\p$ to $\gast\gast$, and
with an increase as a function of $Q^2$, Fig.~\ref{fig:jetrate}a. Again
the detailed studies in Figs.~\ref{fig:jetrate}b--d show that the main effect
in $Q^2$ is not in each individual event class but rather in the mixture of 
the event classes. Thus VMD is similar to $\p\p$, with minor differences from
parton distributions, while GVMD is higher and direct higher still 
(at large $Q^2$).
The DIS rate starts vanishing, since events with parton $\pT > Q$ are put 
in the direct class, but eventually comes up for $Q > \pT$. In that range,
all direct events are required to have jets at lowest order, and only 
fluctuations in the parton shower and hadronization can occasionally stop
jets from being found. 
For $\gast\gast$ events, shown in Figs.~\ref{fig:jetrate}c and~d, 
the single- and double-resolved processes show the behaviour of the 
direct and resolved (VMD plus GVMD) ones, respectively, in $\gast\p$. 
The additional direct class increase slowly with $Q^2$, 
Fig.~\ref{fig:jetrate}c, and dominates the jet rate between the 
single-resolved and DIS dominated regions, peaking at $Q_1^2=1000$~GeV$^2$, 
Fig.~\ref{fig:jetrate}d. It is at this point, for this center of mass 
energy, where the direct matrix elements starts to be suppressed by the large 
photon virtuality.

\begin{figure} [tp]
   \begin{center}
   \mbox{\psfig{figure=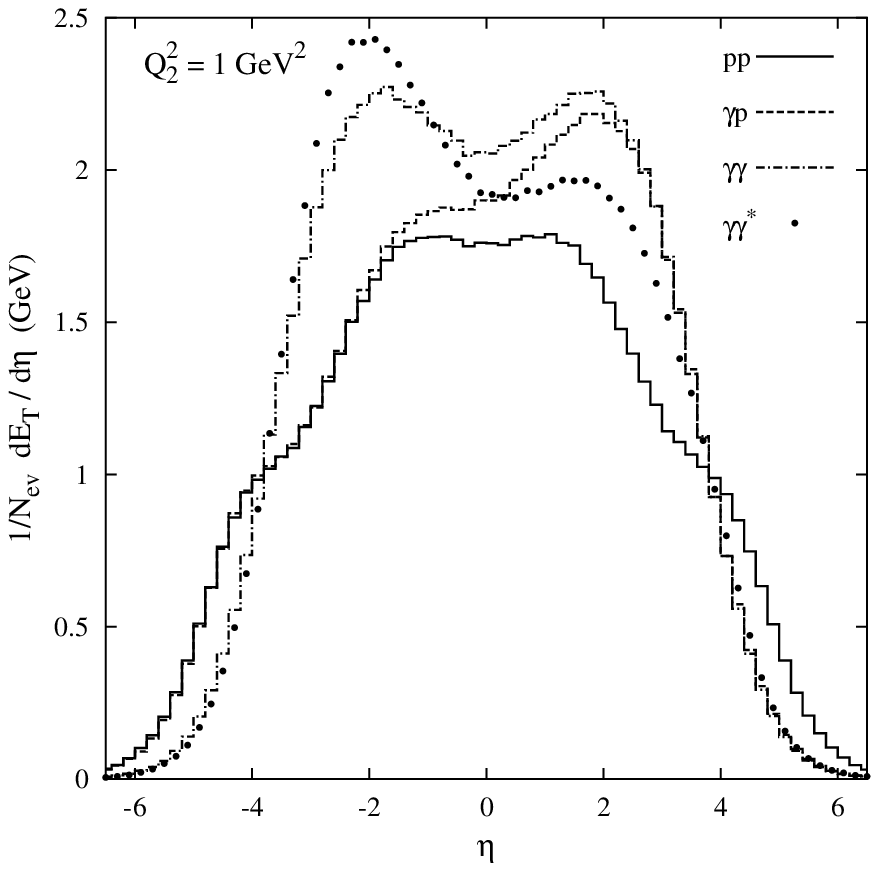,width=105mm}\hspace{-25mm}
	\psfig{figure=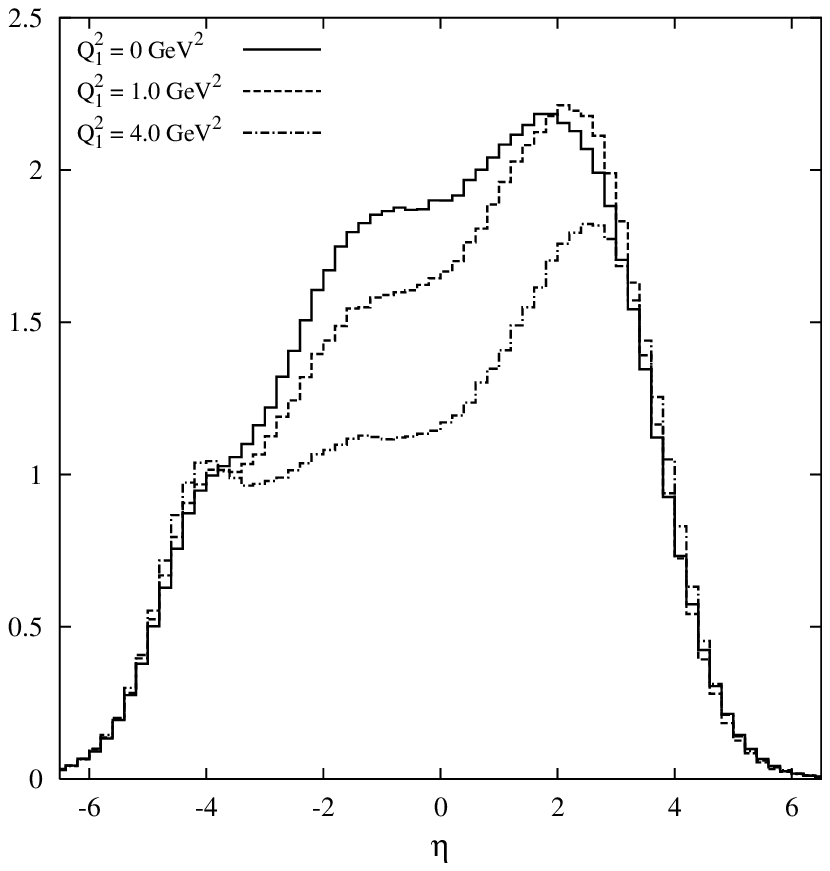,width=105mm}}
   \end{center}
\vspace{-5mm}\hspace{2cm}a)\hspace{8cm}b)
   \begin{center}
   \mbox{\psfig{figure=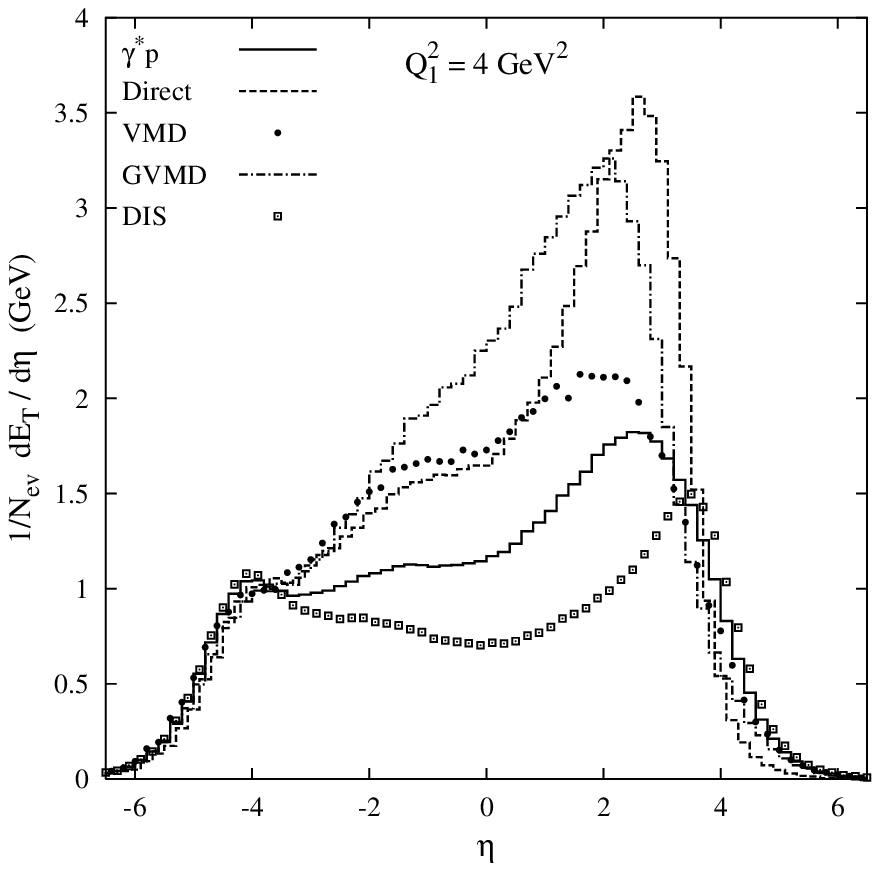,width=105mm}\hspace{-25mm}
	\psfig{figure=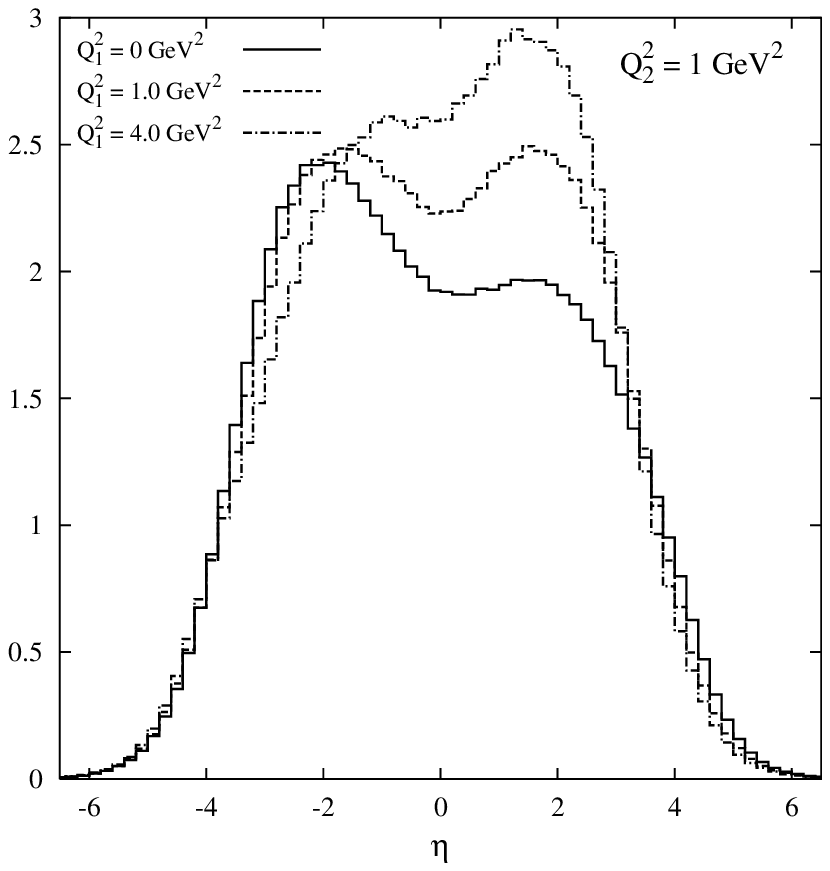,width=105mm}}
   \end{center}
\vspace{-5mm}\hspace{2cm}c)\hspace{8cm}d)\\
\captive{%
a) The transverse energy flow as a function of pseudo-rapidity in the 
center of mass frame of the collision, $W=100~\mr{GeV}$ in all cases.
No diffractive or elastic events are considered.
b) The transverse energy flow as a function of pseudo-rapidity in the $\gast\p$
center of mass frame, normalized to the number of events.
Distributions are shown for photon virtualities of 0, 1.0 and 4.0~GeV$^2$.
c) The different components in $\gast\p$ collisions, normalized to the number 
of events of the respective kind. The photon virtuality is 
$Q_1^2=4~\mr{GeV}^2$.
d) The transverse energy flow as a function of pseudo-rapidity in the 
$\gast\gast$ center of mass frame, normalized to the number of events. 
Distributions are shown for photon virtualities $Q_1^2$ of 0, 1.0 and 
4.0~GeV$^2$, the other photon has the virtuality $Q_2^2=1~\mr{GeV}^2$.
\label{fig:ETflow}}
\end{figure}

\begin{figure} [t]
   \begin{center}
   \mbox{\psfig{figure=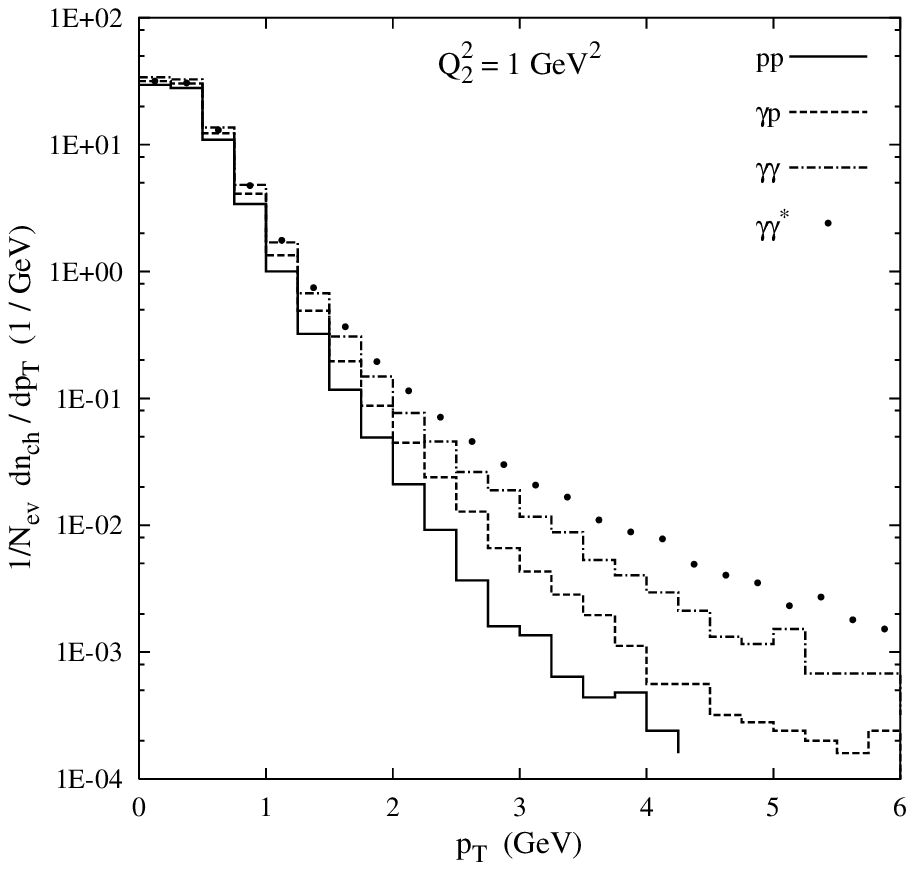,width=105mm}\hspace{-25mm}
	\psfig{figure=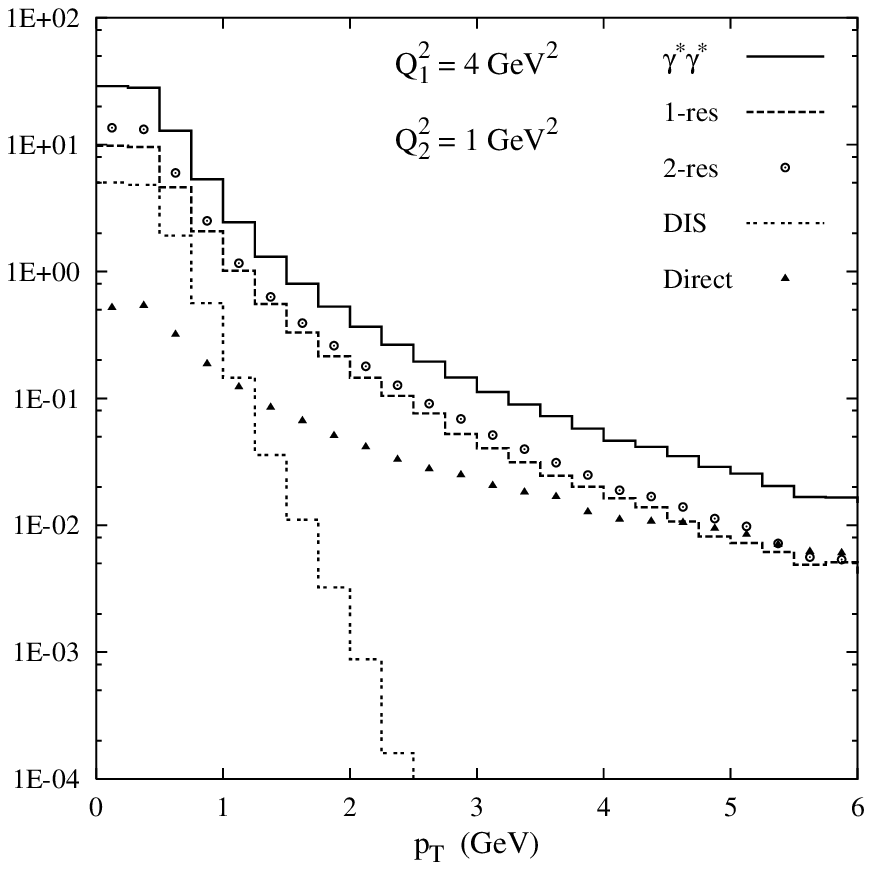,width=105mm}}
   \end{center}
\vspace{-5mm}\hspace{2cm}a)\hspace{8cm}b)\\
\captive{%
a) 
The transverse momentum spectra of charged particles for different collisions, 
normalized to the total number of events. The pseudo--rapidity interval 
is constrained to [0.5,1.5] (corresponding to the $\ga$ direction in case of
$\ga\p$, $\ga\ga$ and $\ga\gast$ collisions). For $\ga\gast$, one photon
is real and the other has a virtuality of $Q_2^2=1~\mr{GeV}^2$.
b) With $Q_1^2=4~\mr{GeV}^2$ and $Q_2^2=1~\mr{GeV}^2$, the $\gast\gast$
distribution is shown for the different components;  single-resolved, 
double-resolved, DIS and direct. The sum of them, add up to the full 
line.
\label{fig:nchpT}}
\end{figure}

The transverse energy flow, as a function of pseudo-rapidity in the center
of mass frame of the collision, is shown in Fig.~\ref{fig:ETflow}. 
(Again, elastic and diffractive events are not considered.) In
Fig.~\ref{fig:ETflow}a all photons are real except for $\ga\gast$ where one 
of them has the virtuality $Q_2^2=1$~GeV$^2$. Here the most interesting 
aspect is to see how the events differ in the photon and the proton 
directions, $\p\p$ being symmetric as well as $\ga\ga$ whereas the $\ga\p$
interpolate between the two with the photon in the $\eta>0$ direction. 
Comparing the $\ga\gast$ with the $\ga\ga$ case, the single-resolved 
processes in $\ga\gast$ has a larger energy flow in the $\gast$ direction 
($\eta<0$) due to the increased jet activity. On the other hand, for 
the double-resolved ones it decreases in both hemispheres but less in the 
$\gast$ direction. Additionally, the DIS processes come in to play but add
on more or less evenly between the two hemispheres at this low virtuality.

In Fig.~\ref{fig:ETflow}b, the transverse energy flow is shown for three
different virtualities in the $\gast\p$ center of mass frame. The major
differences again come from the transition of the events from being 
VMD dominated to being dominated by the DIS processes. 
This is compensated by 
an increasing jet activity in the photon direction, giving the 
rather subtle changes of the total contribution in the photon hemisphere 
which camouflage the big changes of the relative composition of the 
different event classes. 
The different components in $\gast\p$ collisions with $Q_1^2=4$~GeV$^2$ 
are shown in Fig.~\ref{fig:ETflow}c, normalized to the number of events 
of the respective kind. The asymmetry is largest in the direct class, where
all of the photon energy goes into the two high-$\pT$ jets. The typical 
$\pT$ of the jets is not all that high, however, so at least one of the
jets appears at large positive pseudo-rapidities. The VMD and GVMD events are
more symmetric, but the virtual photon parton distributions are harder than 
the corresponding proton one, giving rise to the observed asymmetry. The VMD
events are nearly symmetric for the real photon case, whereas for the 
GVMD ones the asymmetry is still pronounced, as it should be. As expected, 
the largest contribution in the photon hemisphere is in decreasing order 
from DIS, direct, GVMD and VMD events, where the direct and GVMD events 
are of about equal importance. 

A slightly asymmetric energy sharing between the two coloured beam remnants 
in a hadron--hadron kind of collision has been chosen for the results in 
this section. A more asymmetric energy sharing could also well be imagined, 
leaving an uncertainty in the model. For example, the energy flow
in the VMD events of Fig.~\ref{fig:ETflow}c could well decrease by several 
percent for a latter case. 

In Fig.~\ref{fig:ETflow}d, the transverse energy flow for the collision
of two photons is shown. One of the photons is kept at $Q_2^2=1$~GeV$^2$ 
($\eta<0$) and the other has $Q_1^2=0$, 1 and 4~GeV$^2$, all shown separately. 
The $Q_1^2=0$~GeV$^2$ distribution was discussed earlier in comparison 
with two real photons, and is built up by (in order of importance) the 
asymmetric double- and single-resolved contributions together with a small
symmetric DIS contribution --- the direct event class is negligible. 
With both photons having the same virtuality, \mbox{$Q_i^2=1$~GeV$^2$}, 
the above relative importance between the different components still holds 
but now all event classes give a symmetric contribution to the $E_{\perp}$ 
flow. The last case, with two photons at different virtualities, 
the single-resolved processes are responsible for most of the asymmetric 
shape of the total contribution. Now also the direct processes starts to 
be important with a central plateau of two units in pseudo-rapidity. It 
is here comparable with the DIS contribution, about 0.25~GeV at central 
rapidities (normalized to the total number of events). 

The transverse momentum spectra of charged particles become harder when 
going from $\p\p$ to $\ga\p$ to $\ga\gast$, Fig.~\ref{fig:nchpT}, 
which is partly a reflection of the property of the respective parton 
distributions and the pattern seen in the jet rates. 
In $\ga\p$, the VMD processes dominate in the low end of the spectra and the
GVMD and direct processes in the high-$\pT$ tail. With increasing photon 
virtuality, the DIS processes enters but is only of importance at low $\pT$ 
due to the constraint $\pT < Q$. (The parton shower and 
hadronization will cause some particles to be found at $\pT>Q$, however.) 
With a photon virtuality of a few GeV$^2$, a similar spectra is obtained as 
for $\gast\gast$ in Fig.~\ref{fig:nchpT}b, remembering to associate the 
$\ga\p$ direct events with the $\gast\gast$ single-resolved ones, and the 
$\ga\p$ resolved with the $\gast\gast$ double-resolved ones. 

With two photons, the direct processes give a tail out at high $\pT$ that is 
comparable to the single- and double-resolved processes, 
Fig.~\ref{fig:nchpT}b, and contribute to the hardening of the spectra as 
compared to $\p\p$ and $\gamma\p$ collisions.
When both photons have the same virtuality, for example two real photons, 
the DIS processes are as usual absent. Although the mixture of event classes 
differs significantly within a certain kind of collision, the low end of 
the spectra remain approximately the same when the photon virtuality is 
increased. 

\section{Summary and Outlook}

We have in this article tried to outline a scenario that covers `all' 
photon interactions, whether real or spacelike virtual, that produces hadronic
final states. Part of it relies on previous studies, e.g. on jet production
with virtual photons or total cross sections for real photons, but here it 
is eventually combined to give the overall picture. In doing this, 
we attempt to integrate various aspects into a reasonably consistent overall
picture. It would have been nice if a single very economical ansatz for the
photon could be made to cover all relevant phenomenology. We do not 
exclude that this would be possible, although we could also see problems
with such an approach. For instance, it appears likely that there really
is a fundamental separation into a `resolved' and a `direct' 
(or `pointlike') part of the photon, e.g. based on the $x_{\ga}$ 
distributions observed at HERA \cite{xgamma}. Even this separation, of course, 
is only theoretically well-defined to lowest order, and needs to be prescribed
in higher orders of perturbation theory. Nevertheless,
the `all orders' data quite nicely show a separation.

Our model takes such an approach one step further. The resolved part of the
photon is subdivided into VMD and GVMD, depending on whether the resolved 
photon is associated with one of the lowest-lying vector mesons or with
one in the set of not so well known higher-mass resonances. For a virtual 
photon, these states then have a dampened cross section given by dipole form
factors. The direct sector is somewhat more complicated, since the lowest-order
DIS process $\gast \q \to \q$ is not allowed in the limit $Q^2 \to 0$ while
the higher-order direct ones $\gast \q \to \q\g$ and $\gast \g \to \q \qbar$
are. It therefore requires some care to retain only the latter for $Q^2 = 0$
while equating them with the first-order QCD corrections to the DIS process
for large $Q^2$. 

The first test that the mixing makes sense comes from a comparison with 
the total cross sections of $\gast\p$ and $\gast\gast$ interactions. It is 
possible to obtain a reasonable description of all the data, although with
some disagreement for rather small $W$, where our language is not expected
to survive limited phase-space corrections and exclusive final-state effects
anyway. Of course, since we have not even attempted to produce our own sets of 
parton distributions, but taken existing ones, all the possibilities of tuning 
are not exhausted by far. Also the inclusion of the effects of longitudinal
resolved photons remains an area where little is known, and our simpleminded
ansatz could be made more sophisticated, e.g. by the use of parton 
distributions designed to describe the partonic structure of longitudinal
virtual photons \cite{ChylaFl}.
However, the model is so far limited to the SaS~1D set of parton distributions.
An alternative approach with a common resolved class could have been taken,
facilitating the handling of the partonic processes, but probably ending up 
with a more complicated picture for other aspects such as elastic and 
diffractive scattering, multiple parton--parton interactions, etc. and at
the same time losing the flexibility of having the non-perturbative part
of the photon structure separated from the perturbatively calculable one.
It is a subject for further studies, however.

More sophisticated tests come from the study of event shapes. We have in this
article provided some examples how event properties vary between different
initial states and photon virtualities. The examples are mainly chosen as 
simple illustrations; we look forward to more detailed studies by the 
experimental community, based on the code we now provide in the \Py\ event 
generator. Again one cannot expect perfect agreement, but at least an overall 
such, where disagreements hopefully could help provide hints in which 
direction to move for an even better and more complete picture. In this 
sense, our model could be seen as a straw man, although a rather more 
sophisticated such than is often the case.

Given the fairly complex description, we do not expect the model to be
competitive in the high-$Q^2$ region of HERA, say for $Q^2 > 10$~GeV$^2$.
There the simple DIS language provides a powerful starting point, that has
been well studied over the years, with many models developed in extensive
detail \cite{HERAMC}. We are here more interested in the crossover region,
$Q^2 \sim 0.5 - 5$~GeV$^2$, where our model predicts all of the photon
components to have comparable cross sections, and we therefore would expect 
no simple picture to work. In $\gast\gast$ studies at LEP2, it is enough to
have one photon in this region for our approach to offer interesting 
alternatives to other descriptions \cite{LEP2MC}. For smaller $Q^2$,
it smoothly attaches to the existing \Py\ model of real-photon interactions,
while the high-$Q^2$ end has no such correspondence.

There are some areas where we already now know that not enough effort has gone
in to cover the field. One such is the treatment of heavy flavours and in 
particular charm production, both open and closed
(primarily $\Jpsi$), where mass effects are very important. For the lighter 
quarks, like the $\u$ one, the intrinsic `current algebra' mass is negligible
relative to the `constituent' one, that sets the scale e.g. of the $\rho$
 mass. But for a complete charm description, both the impact of the current 
algebra mass scale and of the further QCD-induced confinement mass effects 
have to be considered. Another example of a missing area is that related to 
rapidity gaps, which are already included in the VMD/GVMD sector within a 
traditional Regge language, but currently not in the DIS region, in spite of 
the quite conclusive evidence for this from HERA \cite{rapgapdata}. The two 
aspects come together in the $W$ dependence of the production rate of 
exclusive vector mesons, where data show a steeper rise for $\Jpsi$ than 
lighter mesons, and a steeper rise also at larger $Q^2$ \cite{Qdepeps}. On the 
technical side, initial-state radiation remains to be implemented for the 
lower-$\pT$ DIS process (but is there in the higher-$\pT$ direct ones). 

Therefore the current model should not be considered as the end of the road,
but rather as providing a basic framework that could be further refined.
In the end, data will have to tell whether the approach as such is viable 
or not. In the latter case, a simpler scenario would then be preferable, 
but more likely an even more complex one would be required.

\end{document}